\documentclass[prb,twocolumn,superscriptaddress,float,aps]{revtex4-2}
\usepackage{bm,color,amsmath,amssymb,mathrsfs,latexsym,graphicx,psfrag,tabularx}

\usepackage[hidelinks,colorlinks,linkcolor=blue,
citecolor=blue,urlcolor=blue]{hyperref}

\newcommand{\bs}[1]{\boldsymbol{#1}}

\newcommand{\braket}[2]{\left\langle #1 | #2 \right\rangle}
\newcommand{\bra}[1]{\left\langle#1\right|}
\newcommand{\ket}[1]{\left|#1\right\rangle}

\newcommand{\beq}{\begin{equation}}
\newcommand{\eneq}{\end{equation}}

\newcommand{\addCXL}[1]{{\bf \color{red} #1}}
\newcommand{\addKJ}[1]{{\bf \color{blue} #1}}

\newcommand{\eqnref}[1]{Eq.\,\eqref{#1}}
\newcommand{\figref}[1]{Fig.\,\ref{#1}}

\def\kk{\mathbf{k}}

\usepackage{etoolbox}
\makeatletter
\patchcmd{\@maketitle}{\@author}{\@author\show\@thanks}{}{}
\makeatother

\begin{document}

\title{Engineering Miniband Topology via Band-Folding in Moir\'e Superlattice Materials}


\author{Kaijie Yang}
\thanks{These authors contributed equally.}
\affiliation{Department of Physics, The Pennsylvania State University, University Park, Pennsylvania 16802, USA}
\author{Yunzhe Liu}
\thanks{These authors contributed equally.}
\affiliation{Department of Physics, The Pennsylvania State University, University Park, Pennsylvania 16802, USA}
\author{Frank Schindler}
\affiliation{Blackett Laboratory, Imperial College London, London SW7 2AZ, United Kingdom}
\author{Chao-Xing Liu}
\thanks{Corresponding author.}
\affiliation{Department of Physics, The Pennsylvania State University, University Park, Pennsylvania 16802, USA}

\begin{abstract} 
The emergence of topologically non-trivial flat bands in moir\'e materials provides an opportunity to explore the interplay between topological physics and correlation effects, leading to the recent experimental realization of interacting topological phases, e.g. fractional Chern insulators. In this work, we propose a mechanism of band inversion induced by band-folding from the moir\'e superlattice potential for engineering topological minibands in moir\'e materials. We illustrate this mechanism via two classes of model Hamiltonians, namely the Rashba model and the Bernevig-Hughes-Zhang (BHZ) model, under the moir\'e superlattice potentials. Moir\'e minibands with non-trivial band topology, including $\mathbb Z_2$ number, mirror Chern number and fragile topology, have been found and the topological phase diagram is constructed for these moir\'e models. 
A general theory based on band representations in the mori\'e Brillouin zone is also developed for a generalization of this mechanism to other space groups. Possible experimental realizations of our model Hamiltonian are discussed. 
\end{abstract}

\date{\today}

\maketitle

{\it Introduction -}
Recent experimental realizations of topologically non-trivial flat bands in moir\'e materials, including twisted bilayer graphene\cite{bistritzer2011moire, sharpe2019emergent, serlin2020intrinsic,   zhang2019twisted, bultinck2020mechanism, xie2020nature, xie2021fractional, spanton2018observation, ledwith2020fractional, abouelkomsan2020particle, repellin2020chern, wilhelm2021interplay}, multilayer rhombohedral graphene/hexagonal boron nitride moir\'e heterostructures\cite{chen2020tunable, lu2024fractional, chen2019evidence, chittari2019gate, zhang2019nearly}, twisted monolayer-bilayer graphene\cite{polshyn2020electrical, chen2021electrically}, and transition metal dichalcogenides moir\'e superlattice\cite{li2021quantum, cai2023signatures, park2023observation, zeng2023thermodynamic, xu2023observation, wu2019topological, li2021spontaneous, devakul2021magic},
have led to the successful discovery of a variety of integer and fractional Chern insulator phases. 
Topologically non-trivial flat bands in the existing moir\'e materials all originate from the $K$ point in the atomic Brillouin zone (ABZ), presumably because the effective model around the $K$ point does not respect time reversal $\mathcal T$ so that non-trivial Berry curvature can occur locally around $K$ while the whole system still preserves $\mathcal T$. 
The Coulomb interaction, combined with the local Berry curvature around $K$, can together produce ferromagnetic states that spontaneously break $\mathcal T$ and lead to the quantum anomalous Hall state \cite{chang2013experimental,chang2023colloquium} at integer filling per moir\'e Brillouin zone (MBZ), and the fractional Chern insulator phase at fractional filling. 
In contrast, $\Gamma$ ($\kk=0$) is a $\mathcal T$-invariant momentum in the ABZ and  Kramers' theorem guarantees a spin degeneracy at $\Gamma$. The Kramers' pair of bands generally have opposite Berry curvature, so moir\'e flat bands originating from $\Gamma$ are usually found to be topologically trivial in $\mathcal T$-invariant materials \cite{foutty2023tunable, xian2021realization, angeli2021gamma}. Breaking $\mathcal T$ explicitly via external magnetic fields or magnetic materials has been theoretically proposed to achieve topological flat bands around $\Gamma$ \cite{dong2022dirac, paul2023giant}.

In this work, we introduce a band-folding induced band inversion mechanism to realize topologically non-trivial flat bands around $\Gamma$ in the ABZ for $\mathcal T$-invariant systems under moir\'e superlattice potentials, based on two classes of models, namely the Rashba model\cite{bychkov1984oscillatory} and Bernevig-Hughes-Zhang (BHZ) model\cite{bernevig2006quantum}. The key feature in both classes of models is that the band minimum or maximum is slightly away from $\Gamma$ in the ABZ. 
When the moir\'e length scale introduced by the the moir\'e superlattice potential is comparable to the length scale related to the band minimum or maximum, the band folding from the higher MBZs into the 1st MBZ can lead to a band inversion, thus giving rise to topologically nontrivial flat minibands.

\begin{figure}
\centering
\includegraphics[width=\columnwidth]{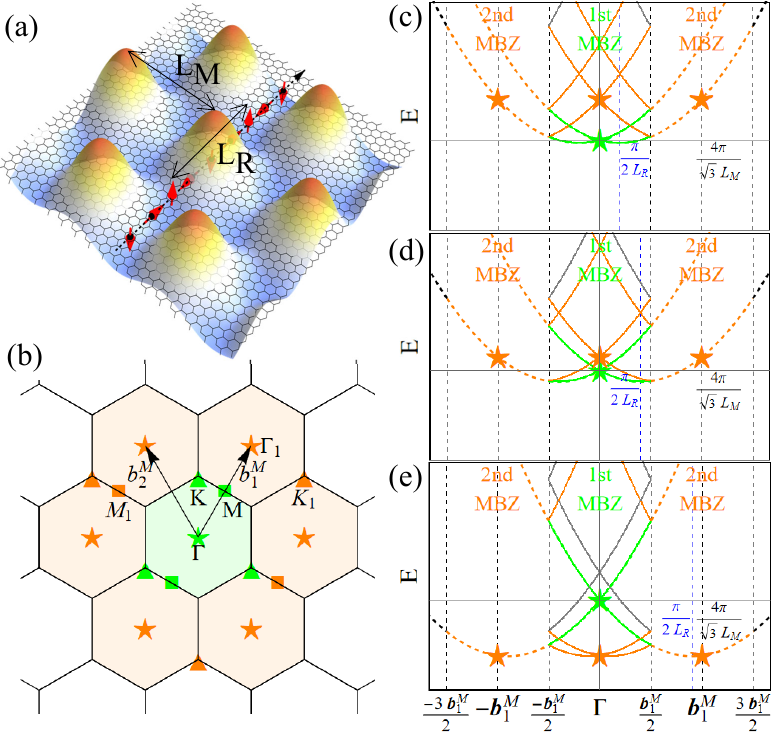}
\caption{
    (a) Schematic figure for moir\'e Rashba systems. An electron on the atomic lattice (black) moves under the moir\'e potential along the black dashed line. Red arrows are electron spins. $L_\text R$ is the spin precession length. $L_\text M$ is the moir\'e unit cell length.
    (b) Extended MBZ with high symmetry points. Green points $\Gamma, M, K$ are located at the green 1st MBZ. Orange points $\Gamma_1, M_1, K_1$ are located at the orange 2nd MBZ.
    (c) - (e) Schematic moir\'e Rashba spectra with increasing Rashba SOC strength or moir\'e unit cell length. The orange dashed lines are unfolded spectra lying in the 2nd MBZ. The green, orange, and gray solid lines are moir\'e minibands folded from the 1st, 2nd and higher MBZ, respectively. The green (orange) star labels the state at $\Gamma$ ($\bs b_1^\text M$). 
}
\label{fig: Rashba schematic}
\end{figure}

The general scenario of the band-folding induced band inversion mechanism for the moir\'e Rashba model is schematically illustrated in \figref{fig: Rashba schematic}, where the topological property of the lowest energy moir\'e miniband is controlled by two length scales: the length scale $L_\text R$ of Rashba spin-orbit-coupling (SOC) that characterizes spin precession\cite{datta1990electronic} and the moir\'e length $L_\text M$, as shown in \figref{fig: Rashba schematic}(a). 
For a weak Rashba SOC with $L_\text R \gg L_\text M$, the band minima are located at $\pi/2L_\text R$ lies in the 1st MBZ (the green region in \figref{fig: Rashba schematic}(b)) and the lowest energy miniband comes from the 1st MBZ in \figref{fig: Rashba schematic}(c). 
When increasing Rashba SOC up to $L_\text R \sim L_\text M $, the band minima move to the boundary between the 1st MBZ and the 2nd MBZ (the orange region in \figref{fig: Rashba schematic}(b)), so that the minibands folded from the 2nd MBZ move down in energy, as shown in \figref{fig: Rashba schematic}(d), and can eventually be lower in energy than the minibands from the 1st MBZ when $L_R \ll L_\text M $, as shown in \figref{fig: Rashba schematic}(e). A band inversion between the mini-bands from the 1st and 2nd MBZs  occurs and leads to the emergence of topologically non-trivial minibands.  
As the Rashba SOC can be tuned via external electric gates, one can electrically engineer band topology in moir\'e Rashba materials. 

The above scenario can also be applied to the BHZ model which is applicable to a class of semiconductor quantum wells (QWs), e.g. HgTe/CdTe QWs\cite{bernevig2006quantum} and InAs/GaSb QWs\cite{liu2008quantum}.  
The valence band maximum in these QWs can be away from $\Gamma$, so that the band-folding induced by a moir\'e superlattice potential can give rise to topological valence minibands (Fig.\ref{fig: bhz phase main}a). We study the topological phase diagram of the highest valence miniband and identify the parameter regimes for topological phases with different mirror Chern numbers and fragile topology. A general formalism based on topological quantum chemistry (TQC) \cite{bradlyn2017topological,cano2021band,elcoro2021magnetic} 
is developed to analyze topological moir\'e minibands for a general space group, which demonstrates that our band-folding mechanism can serve as an efficient approach to induce non-trivial band topology in a class of moir\'e materials. 

\begin{figure}
\centering
\includegraphics[width=\columnwidth]{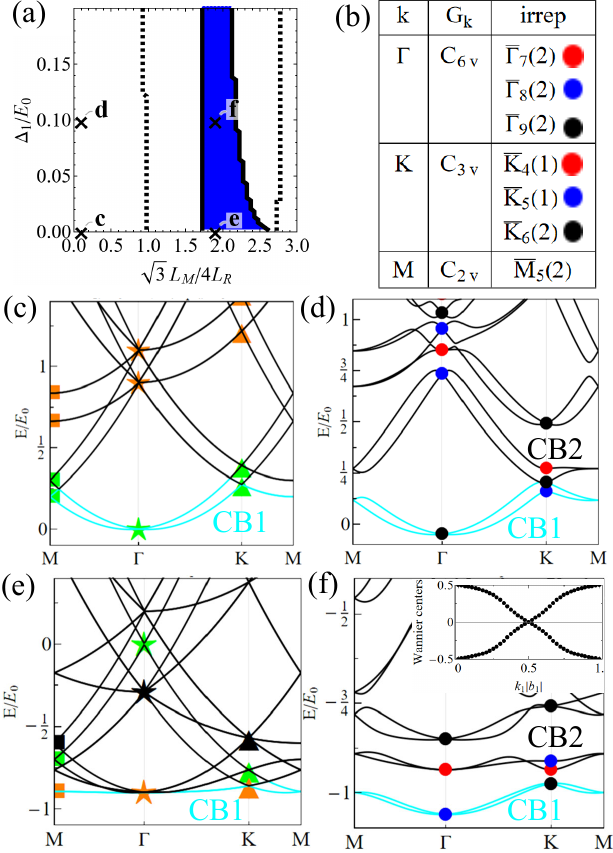}
\caption{
    (a) Topological phase diagram of the lowest-energy conduction band CB1. Blue regions are topological insulator phases with nontrivial $\mathbb{Z}_2$. 
    White regions are semi-metal phases. 
    (b) High symmetry points $\Gamma, M, K$, wavevector group $G_{\bs k}$, and symmetry irreps with dimensions in the parentheses. Colored dots represent irreps in spectra (d)(f).
   Black dashed lines in (a) denote the exchange of irreps $\bar \Gamma_8$ and $\bar \Gamma_9$. 
   Black solid lines denote the exchange of symmetry irreps at $K$ between $\bar K_6$ and $\bar K_4 $,  or between $\bar K_6$ and $ \bar K_5$.
    (c)-(f) spectra with parameters at points c-f in (a), respectively. Colored dots in (c)(e) represent states at extended MBZ points in \ref{fig: Rashba schematic}(b). Green (orange) color represents the 1st (2nd) MBZ. Insets in (f) show Wannier center flow for CB1.
}
\label{fig: Rashba topological phase diagram}
\end{figure}
\begin{figure}
\centering
\includegraphics[width=1\columnwidth]{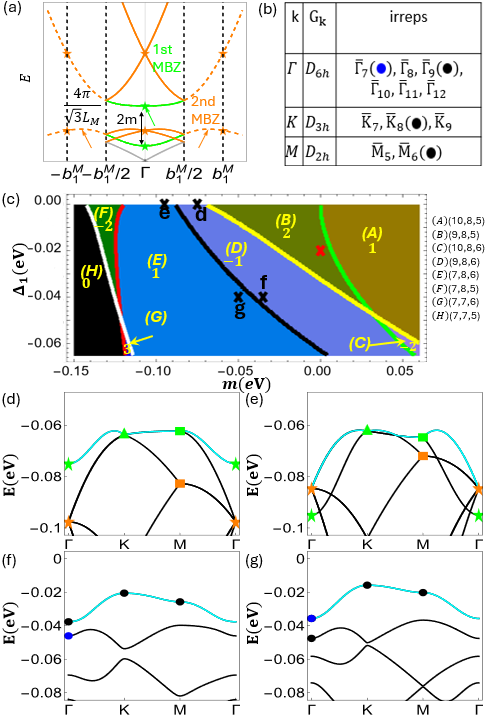}
\caption{ 
(a) Schematic band folding in the moir\'e BHZ system. The orange dashed lines are unfolded spectra lying in the 2nd MBZ. The green (orange) solid lines are moir\'e minibands folded from the 1st (2nd) MBZ. The gray solid lines are minibands from higher MBZ. The green (orange) star labels the state at $\Gamma$ ($|\bs b_1^\text M|$).
$2m$ labels the gap at $\Gamma$. The orange arrow points to the maximum of the valence band.
(b)  Wavevector group $G_k$ of high symmetry points $\Gamma, M, K$ for space group p6/mmm. All symmetry irreps are 2D. Colored dots are representation of irreps in spectra in (f) and (g).
(c) Topological phase diagram of VB1. 
Different color regions (A)-(H) distinguish irreps at high symmetry points, labeled by ($i,j,k$), which is short for ($\bar{\Gamma}_i,\bar{ K}_j,\bar{M}_k$), of VB1. 
The number shows the mirror Chern number $C_{\mathcal M}$. 
The colored solid lines between different regions identify irreps interchanges at high symmetry points. Band dispersion for the parameters labeled by the red cross is shown in the Fig.\ref{fig:band main}a.
(d)-(g) Band dispersion with the parameters labeled by the black crosses d-g in (c). The VB1 is highlighted by cyan. }
\label{fig: bhz phase main}
\end{figure}

{\it Moir\'e Rashba systems -}
Our first example to illustrate this mechanism is the moir\'e Rashba system, which is schematically shown in \figref{fig: Rashba schematic}(a) with the model Hamiltonian
  $  H = H_\text R + H _ \text M $,
where $H_\text R$ is the Rashba model Hamiltonian
\begin{equation} \label{eq: rashba ham}
    H_\text R (\bs k) = a k^2 \sigma_0 + \lambda (k_y \sigma_x - k_x \sigma_y)
\end{equation}
where $\sigma_{0}$ is identity matrix, $\sigma_{x,y,z}$ are Pauli matrices for spin operators, $a > 0$ is the parameter characterizing the effective mass, and $\lambda > 0$ is the Rashba SOC strength that induces spin precession with the length $L_\text R = \pi a /\lambda$. 
We consider a hexagonal moir\'e superlattice potential 
\begin{equation} \label{eq: moire potential main}
    H _ \text M (\bs r) = \Delta_1 \sum_{\bs g}  e^{i \bs g \cdot \bs r} \sigma_0, 
\end{equation}
where $\bs g$ labels the moir\'e reciprocal lattice vectors and $\Delta_1$ denotes the moir\'e potential strength. We consider $\bs g$ up to the first shell in the MBZ, $\bs g = \pm \bs b_1^\text M, \pm \bs b_2^\text M, \pm(\bs b_1^\text M -\bs b_2^\text M)$, as marked by orange stars in \figref{fig: Rashba schematic}(b), where $\bs b_1^\text M = 4\pi/\sqrt 3 L_\text M(1/2, \sqrt 3/2), \bs b_2^\text M = 4\pi/\sqrt 3 L_\text M(-1/2, \sqrt 3/2)$ are the primitive moir\'e reciprocal lattice vectors and $L_\text M$ is the moir\'e unit cell length.
$E_0 = a \vert \bs b_1^\text M \vert^2$ sets the energy unit. 
We consider the space group $P6mm$ that corresponds to the point group $C_{6v}$\cite{elcoro2017double} for  moir\'e Rashba materials. To analyze band topology of this model using TQC \cite{bradlyn2017topological,cano2021band,elcoro2021magnetic}, we need to identify the possible irreducible representations (irreps) of the minibands at all high symmetry momenta, including $\Gamma, M$ and $K$, in the MBZ. The wave vector groups $G_{\bs k}$ and the double group irreps at these momenta are listed in \figref{fig: Rashba topological phase diagram}b. 
We find the low-energy moir\'e minibands can be characterized by $\bar \Gamma_7, \bar \Gamma_8, \bar \Gamma_9$ at $\Gamma$, which are all 2D irreps due to $\mathcal T$, two 1D irreps $\bar K_4, \bar K_5$ and one 2D irrep $\bar K_6$ at $K$, and only one 2D irrep $\bar M_5$ at $M$. More details about the irreps are shown in the character tables in SM Sec.~\ref{sec: moire rashba tqc}.

Tuning the Rashba parameter $\lambda$ and moir\'e length $L_M$ to move the Rashba band minima across the MBZ boundary can lead to non-trivial bulk topology via the band folding mechanism.
The topological phase diagram of the lowest energy minibands, denoted as CB1, as a function of $\sqrt 3 L_\text M/4L_\text R$ and $\Delta_1/E_0$ is shown in \figref{fig: Rashba topological phase diagram}a. 
When $L_\text R \gg  \sqrt 3 L_\text M/ 4$, the CB1 has a band touching with higher energy minibands, so it is in the semimetal phase.
When $L_\text R \sim \sqrt 3 L_\text M/ 8 < L_\text M  \sqrt 3/ 4$, we find a blue region in which the CB1 carries nontrivial $\mathbb Z_2$ topology.
\figref{fig: Rashba topological phase diagram}c-f show the energy spectrum of minibands for the parameters corresponding to the points c-f in \figref{fig: Rashba topological phase diagram}a, and illustrate the occurrence of a band inversion through the exchange of irreps at high symmetry points $\Gamma, K$.  
\figref{fig: Rashba topological phase diagram}c depicts the energy dispersion of minibands for $ \sqrt 3 L_\text M / 4L_\text R = 0.1$ and $\Delta_1=0$ (the point c in \figref{fig: Rashba topological phase diagram}a), where the states of CB1 at $\Gamma, K, M$ all come from the 1st MBZ shown by the green markers in \figref{fig: Rashba schematic}b.
\figref{fig: Rashba topological phase diagram}d shows the energy dispersion for the same $ \sqrt 3 L_\text M / 4 L_\text R$ but a non-zero moir\'e potential $\Delta_1=0.1 E_0$. Two spin degenerate states of CB1 at $\Gamma$ belong to the $\bar \Gamma_9$ irrep, represented by black dots illustrated in \figref{fig: Rashba topological phase diagram}b. At $K$, two spin states of CB1 are split. The lower energy state of CB1 belongs to the 1D irrep $\bar K_5$ while the higher energy state of CB1 belongs to the 2D irrep $\bar K_6$ and is degenerate with another higher energy miniband of CB2, resulting in a semimetal phase for point d in \figref{fig: Rashba topological phase diagram}a. 
In \figref{fig: Rashba topological phase diagram}e with $ \sqrt 3 L_\text M / 4L_\text R = 1.9$ and $\Delta_1=0$, the CB1 at $\Gamma$, $M$ and $K$ are all from the 2nd MBZ, as shown by the orange markers in \figref{fig: Rashba schematic}b. 
Increasing the moir\'e potential to $\Delta_1=0.1 E_0$, we notice that the irreps of CB1 at $\Gamma$ and $K$ belong to $\bar \Gamma_8$ and $\bar K_6$, respectively, both of which are 2D irreps (SM Sec.II.A). 
Thus, CB1 are separated from CB2 by a mini-gap and become isolated moir\'e minibands. 
The Wannier center flow \cite{yu2011equivalent} in the insets of \figref{fig: Rashba topological phase diagram}f demonstrates nontrivial $\mathbb Z_2$ topology of CB1. 
In \figref{fig: Rashba topological phase diagram}a, the dashed line around $L_\text R \sim L_\text M \sqrt 3/4$ labels the interchange of $\bar \Gamma_9$ from the 1st MBZ and $\bar \Gamma_8$ from the 2nd MBZ for CB1 at $\Gamma$, while the solid line that separates white and blue regions represents the interchange of the $\bar K_6$ and $\bar K_5$ minibands at $K$. 
The interchange between the minibands from different MBZs is the origin of non-trivial topology for  CB1.  %

\begin{figure}
\centering
\includegraphics[width=1\columnwidth]{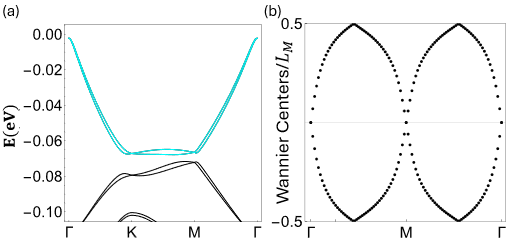 }
\caption{
\label{fig:band main}
(a) Band dispersion with Rashaba SOC strength $\lambda$=0.02 nm$\cdot$eV 
and parameters labeled by the red cross in Fig.~\ref{fig: bhz phase main}c. (b) Wannier center flow for VB1. The Wannier center flow still reveals a double winding feature when Rashba SOC is included. }
\label{fig4}
\end{figure}

{\it Moir\'e BHZ systems -}
Our second example is the BHZ model that has been successfully applied to HgTe QWs \cite{rothe2010fingerprint}, in which the early literature has revealed the valence band maximum away from $\Gamma$ in the inverted band regime \cite{ortner2002valence,novik2005band}.
The Hamiltonian of the BHZ model is described in SM Sec.III.A, and the key parameter is the gap parameter, denoted as $m$ in \figref{fig: bhz phase main}a, which controls the normal or inverted band structure, depending on its sign. \figref{fig: bhz phase main}a shows the schematic unfolded band spectrum of the BHZ model with the valence band maxima located around $k_{max}\sim \sqrt{|m/B|}$
away from $\Gamma$, where $B$ is the coefficient of $k^2$ term in the BHZ model, as depicted by the orange arrow in \figref{fig: bhz phase main}a, which can be achieved for a negative $m$ (SM Sec.III.A).
The moir\'e potential has the same form as \eqnref{eq: moire potential main} with the potential strength $\Delta_1$. 
When the moir\'e reciprocal lattice vector $\vert \bs b_1^\text M \vert$ 
is comparable to the valence band maxima $k_{max}$, we expect the band inversion due to band-folding can occur. We consider the space group for the moir\'e BHZ system to be $P6/mmm$ corresponding to the point group $D_{6h}$\cite{elcoro2017double}. Due to the combined symmetry of $\mathcal T$ and inversion $\mathcal I$, all irreps are 2D so that all minibands are doubly degenerate for spin, which facilitates the existence of isolated minibands, in sharp contrast with spin-split bands in the Rashba model due to the absence of inversion $\mathcal{I}$. At inversion-symmetric momenta $\Gamma$ and $M$, mini-bands can be characterized by $\bar \Gamma_7, \bar \Gamma_8, \bar \Gamma_9$ and $\bar M_5$ with even parities under $\mathcal I$, and $\bar \Gamma_{10}, \bar \Gamma_{11}, \bar \Gamma_{12}$ and $\bar M_6$ with odd parities. At $K$, three 2D irreps $\bar K_7, \bar K_8, \bar K_9$ can appear for minibands. All these symmetry properties of minibands are summarized in Fig. \ref{fig: bhz phase main}b.




Due to higher symmetry and non-trivial band topology in the unfolded BHZ model, a rich topological phase diagram emerges for the mori\'e BHZ model, as compared to the mori\'e Rashba model. Here we focus on the top valence minibands, denoted as VB1, in the negative $\Delta_1$ regime, in which the VB1 feels a moir\'e potential of honeycomb lattice. Due to the existence of the ouf-of-plane mirror symmetry $\mathcal M_z$ \cite{teo2008surface, ando2015topological}, the topological phase diagram can be characterized by 
the mirror Chern number $C_{\mathcal M}$. 
The mirror Chern number $C_{\mathcal M}$ and the irreps at high symmetry momenta ($\Gamma, K, M$) for the VB1 as a function of $m$ and $\Delta_1$ are summarized in \figref{fig: bhz phase main}c, in which multiple topological phase transition lines are found and have different physical origins.
For a positive $m$ and a small value of $|\Delta_1|$, the valence band maximum of the unfolded BHZ model is at $\Gamma$ and thus the phase transitions for VB1 (green and yellow lines in \figref{fig: bhz phase main}c) only involve the states within the 1st MBZ. The green transition line between the phases A and B corresponds to the band inversion of the unfolded BHZ model at $\Gamma$ and changes $C_{\mathcal M}$ by 1, while the yellow transition line between the phases B (A) and D (C) corresponds to the band inversion at three equivalent $M$ at the boundary of the 1st MBZ and thus changes $C_{\mathcal M}$ by 3 (SM Sec.III.B). Further reducing $m$ towards the black transition line in \figref{fig: bhz phase main}c, the valence band maximum moves towards the 2nd MBZ, as depicted in  \figref{fig: bhz phase main}d and e at $\Delta_1=0$, and the band-folding mechanism starts working. 
Across the black transition line, the energy of the $\bar{\Gamma}_7$ state at $\Gamma$ in the 2nd MBZ becomes lower than that of the $\bar{\Gamma}_9$ state in the 1st MBZ, so a band inversion occurs for the VB1 due to the exchange of the $\bar{\Gamma}_9$ and $\bar{\Gamma}_7$ irrep states, as shown in \figref{fig: bhz phase main}f and g. We also find that this transition has the form of a quadratic band touching at the transition point (SM Sec.III.B) and changes $C_{\mathcal M}$ by 2. Consequently, the $C_{\mathcal M}$ of VB1 changes from $C_{\mathcal M}=-1$ in region D to $C_{\mathcal M}=1$ in region E (\figref{fig: bhz phase main}f and g). With further reducing $m$, the red transition line is for the band inversion for the minibands at $M$ of the 2nd MBZ, which varies $C_{\mathcal M}$ by 3, while the white line is for the transition of the minibands at $K$ in the 2nd MBZ that have have lower energy than those at $K$ in the 1st MBZ. As there are two different $K$s in the MBZ, the mirror Chern number of VB1 changes by 2 across the white line, and we find $C_{\mathcal M}=-2$ in region F and $C_{\mathcal M}=0$ in region H.

We next point out unique aspects of the minibands for the moir\'e BHZ model, which are absent for the moir\'e Rashba model. We first notice the VB1 with $C_{\mathcal{M}}=\pm 2$ in the B, C and F regions. These phases possess two copies of helical edge modes at the boundary. Previous theoretical studies \cite{bi2017bilayer,zhang2017fingerprints,zhang2016interacting}
proposed strong interactions in the $C_{\mathcal{M}}=\pm 2$ phase can give rise to a bosonic symmetry protected topological phase, which is characterized by a bosonic mode at the boundary instead of a fermionic mode. When mirror symmetry is broken by applying an out-of-plane electric field via gate voltages \cite{winkler2003spin}, $C_{\mathcal{M}}$ loses its physical meaning and the corresponding helical edge modes are expected to gap out. The band dispersion and Wannier center flow in this case are shown in \ref{fig4}a and b as an example, respectively. Strikingly, the Wannier center flow of VB1 shown in \ref{fig4}b still has a  winding number $w=2$, 
independent of mirror symmetry breaking. This nontrivial double winding of Wannier centers indicate fragile topology \cite{po2018fragile, alexandradinata2014wilson, bradlyn2019disconnected, cano2018topology, wieder2018axion,bouhon2019wilson,wieder2020strong}, protected by $\mathcal C_{2z} \mathcal T$, where $\mathcal C_{2z}$ is the two-fold rotation symmetry. The nature of fragile topology can be revealed by including additional trivial bands to gap out the double winding of Wanner centers, as discussed SM Sec.III.C. 
In addition, the topologically nontrivial minibands in region E are relatively flat with its miniband width smaller than 2.5 meV and minigap around 20 meV, potentially supporting a fractional Chern insulator phase, which was very recently discussed in Ref.\cite{tan2024designing}.

{\it Conclusion and discussion -}
In conclusion, the band-folding mechanism can give rise to band inversion of moir\'e minibands in different extended MBZs and induces nontrivial $\mathbb Z_2$ topology, fragile topology, mirror Chern topology and topological flat minibands in the Rashba and BHZ systems under moir\'e superlattice potentials.
The band-folding mechanism for topological minibands can be generalized to other moir\'e materials with different space group symmetries. The SM Sec.~\ref{sec:moire TQC} describes a general approach based on the TQC method\cite{bradlyn2017topological,elcoro2017double,vergniory2017graph} to classify the irreps of minibands by combining the group theory analysis of band-folding mechanism with the perturbation theory in the weak moir\'e potential limit. Applying this method to the moir\'e Rashba and BHZ models is also discussed in SM Sec. \ref{sec:MoireRashba} and \ref{sec:moire BHZ}, respectively.

A class of semiconductor heterostructures with large Rashba SOC and high electron mobility is identified n the SM Sec.II.B for possible material realizations of moir\'e Rashba systems. The BHZ model can be realized by HgTe/CdTe QWs\cite{bernevig2006quantum, konig2007quantum}, InAs/GaSb QWs\cite{liu2008quantum, du2015robust}, and monolayer 1T$'$-WTe$_2$\cite{tang2017quantum}. The superlattice potential could be created by moir\'e 2D insulating materials, such as boron nitride or transition metal dichalcogenides\cite{yasuda2021stacking, zhao2021universal,kim2024electrostatic,woods2021charge, wang2022interfacial,xu2021creation}, or by the patterned hole array in dielectric substrate materials to form a superlattice potential\cite{forsythe2018band, barcons2022engineering}. The great tunability of moir\'e structure and the potential strong Coulomb interaction of moir\'e minibands may allow us to realize bosonic SPT phase \cite{zhang2016interacting, bi2017bilayer} and $\mathcal C_{2z} \mathcal T$-protected $w=2$ fragile topology \cite{schindler2019fractional}, which have not been realized in atomic crystals so far.




{\it Acknowledgement -}
We thank Liang Fu, Jainendra K Jain, Zhen Bi, Jiabin Yu, Lujin Min, Ke Huang for helpful discussion. K.J.Y., Y.Z.L. and C.X.L. acknowledge the support from the NSF through The Pennsylvania State University Materials Research Science and Engineering Center [DMR-2011839]. C.-X.L. also acknowledge support from NSF grant via the grant number DMR-2241327 and also the grant NSF PHY-1748958 to the Kavli Institute for Theoretical Physics (KITP).

%

\clearpage

\pagebreak
\widetext
\begin{center}
	{\large{\bf Supplementary Materials}}
\end{center}

\setcounter{equation}{0}
\setcounter{figure}{0}
\setcounter{secnumdepth}{2}
\setcounter{page}{1}
\setcounter{section}{0}
\renewcommand{\theequation}{S\arabic{equation}}
\renewcommand{\thefigure}{S\arabic{figure}}
\renewcommand{\thetable}{S\arabic{table}}

In this Supplementary Material, we will first describe our general approach to identify topological property of moir\'e minibands based on topological quantum chemistry (TQC) method \cite{bradlyn2017topological,elcoro2017double,vergniory2017graph} in Sec. \ref{sec:moire TQC}. Then, we will apply this approach to moir\'e Rashba systems in Sec. \ref{sec:MoireRashba} and moir\'e Bernevig-Hughes-Zhang (BHZ) model in Sec. \ref{sec:moire BHZ}. 

\section{General formalism for Topological properties of moir\'e systems based on topological quantum chemistry}\label{sec:moire TQC}

\begin{figure}
\centering
\includegraphics[width=\columnwidth]{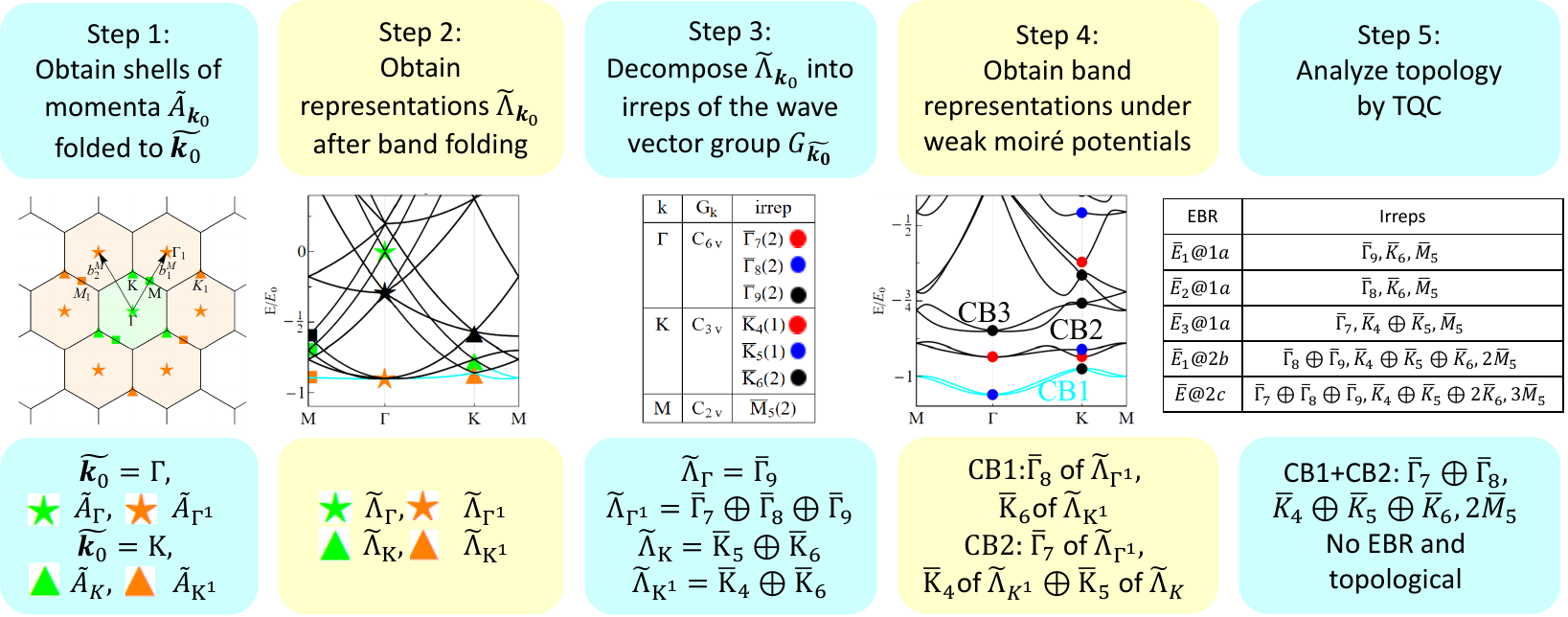}
\caption{
    The flow chart of moir\'e TQC with the moir\'e Rashba systems as an example. The first row are the goals for each step. The second row are figures for the moir\'e Rashba systems adapted from \figref{fig: Rashba schematic} and \figref{fig: Rashba topological phase diagram} in the main text. The table under the step 5 lists irreps at high symmetry points for all EBR in the space group $P6mm$. The third row is the results from each step.
}
\label{fig: moire TQC}
\end{figure}

In this section, we will develop a general approach to derive band representations of moir\'e minibands at high symmetry momenta under a weak moir\'e superlattice potential to analyze miniband topology based on the theoretical framework of topological quantum chemistry (TQC)\cite{bradlyn2017topological,elcoro2017double,vergniory2017graph}.
The whole procedure can be implemented in five steps and a flow chart, using the moir\'e Rashba model as an example, is shown in Fig.\ref{fig: moire TQC}. 



Step 1: we define the moir\'e reciprocal lattice vectors in atomic Brillouin zone (ABZ) and label the shells of momenta in the ABZ that are folded into the same momentum in the 1st moir\'e Brillouin zone (MBZ). A tight-binding Hamiltonian or an effective Hamiltonian, $H_0(\bs k)$, describes the electronic band structure of an atomic crystal with its crystal symmetry described by the space group, denoted by $G^0$, and the crystal momentum $\bs k$ defined in the ABZ of $G^0$. For the effective model that is expanded around certain momentum in the ABZ, we can choose a crystal symmetry group that is compatible with the full symmetry property of the effective model. As an example, we choose the $P6mm$ group for the Rashba model and the $P6/mmm$ group for the BHZ model in this work. It should be emphasized that the same effective model can be applied to crystal materials with different space groups. At a certain momentum $\bs{k_0}$ in the ABZ, the wave vector group (\emph{i.e.}, the little group that leaves this wave vector invariant) is denoted as $G_{\bs{k_0}}^0$.
The moir\'e superlattice potential $H_M$ can be described by a different space group, denoted as $G^\text M$, so the space group for the whole moir\'e system is described by the group $G = G^\text M \cap G^0$ that can be used to define the MBZ. We denote $\bs g_0$ and $\bs g$ as the reciprocal lattice vectors for the space group $G^0$ and $G$, respectively. For any momentum $\bs k_0$ in the ABZ, one can always find a momentum $\tilde {\bs k}_0$ in the 1st MBZ so that $\bs k_0$ and $\tilde {\bs k}_0$ are connected by a moir\'e reciprocal lattice vector $\bs g$. The set of all momenta in the ABZ that are folded to the same $\tilde {\bs k}_0$ is defined as $A_{\tilde {\bs k}_0} =\{\bs k_0 \vert \bs k_0=\tilde {\bs k}_0 + \bs g\}$. The wave vector group at $\tilde {\bs k}_0$ for the space group $G$ is denoted as $G_{\tilde {\bs k}_0}$, which includes all the point group symmetry operators in $G$ that transforms $\tilde {\bs k}_0$ into a momentum ${\bs k}_0 \in A_{\tilde {\bs k}_0}$. 
\begin{equation}
    G_{\tilde {\bs k}_0} = \{ \mathcal S \vert   \mathcal S\tilde{\bs k}_0\in A_{\tilde {\bs k}_0} \And \mathcal S\in O(3) \cap G\},
\end{equation}
where $O(3)$ is the three dimensional orthogonal group. $G_{\bs{k_0}}^0$ and $G_{\tilde {\bs k}_0}$ can contain different symmetry operators for ${\bs k}_0 \in A_{\tilde {\bs k}_0}$. The point group of $G$ is the wave vector group $G_{\Gamma}$, where $\Gamma$ is the origin of the MBZ. Using the wave vector group $G_{\tilde {\bs k}_0}$, the momentum shell, denoted as $\tilde A_{\bs k_0}$ for any momentum ${\bs k}_0 \in A_{\tilde {\bs k}_0}$, can be defined by applying all symmetry operators $\mathcal S \in G_{\tilde {\bs{k}}_0}$ to $\bs k_0$ and then collecting distinct momenta, namely
\begin{equation}\label{eq:shell of momenta}
     \tilde A_{\bs k_0} = \{ {\bs k} \vert  \exists \mathcal S\in G_{\tilde{\bs k}_0}, \bs k = \mathcal S \bs k_0 \}.
\end{equation}
As $\mathcal S \in O(3)$, all the momenta in one momentum shell $\tilde A_{\bs k_0}$ must have the same amplitude, and thus we can define the momentum shell amplitude as $|\tilde A_{\bs k_0}|=|\bs k_0|$. 
We arrange the momentum shells according to the ascending order of the momentum shell amplitude $|\tilde A_{\bs k_0}|$, and dub them as the 1st momentum shell, the 2nd momentum shell and so on. 
As an example, we consider the hexagonal MBZ in "Step 1" of Fig.\ref{fig: moire TQC}, in which one can find only one momentum (green star) in $\tilde A_{\Gamma}$, six momenta (orange stars) in $\tilde A_{\Gamma^1}$, three momenta (green triangles) in $\tilde A_{K}$ and three momenta (orange triangles) in $\tilde A_{K^1}$. 

Step 2: the eigen-energy bands of $H_0({\bs k}_0)$ at the momentum ${\bs k}_0\in A_{\tilde {\bs k}_0}$ are folded into the momentum $\tilde {\bs k}_0$ in the 1st MBZ and the representation of the folded bands at $\tilde {\bs k}_0$ can be constructed under zero moir\'e potential $H_M=0$. 
Here we consider a set of degenerate atomic energy bands, denoted as $\ket {\Lambda_{\bs k_0}^0, n}$, of $H_0$ at $\bs k_0$ in the ABZ with the eigen-energy $E^{\Lambda}_{\bs k_0}$. We further assume this set of basis wavefunctions belong to the irreducible representation (irrep) $\Lambda_{\bs k_0}^0$ of the group $ G_{\tilde{\bs k}_0} \cap G_{\bs k_0}^0$, which is a subgroup of both $G_{\tilde{\bs k}_0}$ and $G_{\bs k_0}^0$. 
Here $n=1,2,...,N[\Lambda_{\bs k_0}^0]$ and $N[\Lambda_{\bs k_0}^0]$ labels the dimension of the irrep $\Lambda_{\bs k_0}^0$. For a symmetry operator $\mathcal S_0 \in G_{\tilde{\bs k}_0} \cap G_{\bs k_0}^0$, its representation matrix $\mathcal{D}_{\bs k_0}^0(\mathcal S_0 )$ and the corresponding character $\chi^0_{\bs k_0}(\mathcal S_0)$ are given by 
\beq \label{eq:repremat_Chara_1}
\mathcal S_0 \ket {\Lambda_{\bs k_0}^0, n} = \sum_{n'} \ket {\Lambda_{\bs k_0}^0, n' } \mathcal{D}^0_{\bs k_0, n' n}(\mathcal S_0 ), \quad \quad \chi^0_{\bs k_0}(\mathcal S_0) = \text{Tr}[\mathcal{D}^0_{\bs k_0}(\mathcal S_0 )],\eneq 
as exemplified in Tab.~\ref{tab: step 2 character table}(a). Next we consider the electronic bands at another different momentum $\bs k'= \mathcal S \bs k_0 \in \tilde A_{\bs k_0}$ with $\mathcal S \in G_{\tilde{\bs k}_0}$ and $\mathcal S \notin G_{\bs k_0}^0$ and construct the corresponding basis wavefunctions by $\ket {\Lambda_{\bs k'}^0, n} = \mathcal S \ket {\Lambda_{\bs k_0}^0, n}$, which are eigen-states of $H_0$ at $\bs k'$ with the same eigen-energy, 
\begin{equation}
   H_0(\bs k') \ket {\Lambda_{\bs k'}^0, n} = \mathcal S H_0(\bs k_0) \mathcal S^{-1} \mathcal S \ket {\Lambda_{\bs k_0}^0, n} = E^{\Lambda}_{\bs k_0} \ket {\Lambda_{\bs k'}^0, n}.
\end{equation}
As the basis wavefunctions at $\bs k_0$ and $\bs k'$ are connected by symmetry operator, $\ket {\Lambda_{\bs k'}^0, n'}$ must belong to the same irrep, denoted as $\Lambda^0_{\bs k'}$, as $\ket {\Lambda_{\bs k_0}^0, n}$. The characters of $\Lambda^0_{\bs k'}$ are summarized in Tab.~\ref{tab: step 2 character table}(b). The wave vector group $G_{\bs k'}^0$ at $\bs k'$ is isomorphic to $G_{\bs k_0}^0$ at $\bs k_0$ because for any $\mathcal S_0 \in G_{\bs k_0}^0$, the symmetry operator $\mathcal S'_0 = \mathcal S \mathcal S_0 \mathcal S^{-1}$ is an element in the wave vector group $G_{\bs k'}^0$ at $\bs k'$. The character of the symmetry operator $\mathcal S'_0$ in $\Lambda^0_{\bs k'}$, denoted as $\chi^0_{\bs k'}(\mathcal S_0')$, should remain the same, namely $\chi^0_{\bs k'}(\mathcal S_0') = \chi^0_{\bs k}(\mathcal S_0)$, because their representation matrices are the same, 
\begin{equation}\label{eq:band representation after MBZ folding}
    \mathcal{D}^0_{\bs k', n_1 n_2}(\mathcal S'_0) = \bra{\Lambda_{\bs k'}^0, n_1} \mathcal S_0' \ket {\Lambda_{\bs k'}^0, n_2} = \bra{\Lambda_{\bs k_0}^0, n_1} \mathcal S^{-1} \mathcal S_0' \mathcal S \ket {\Lambda_{\bs k_0}^0, n_2} = \bra{\Lambda_{\bs k_0}^0, n_1} \mathcal S_0 \ket {\Lambda_{\bs k_0}^0, n_2} = \mathcal{D}^0_{\bs k_0, n_1 n_2}(\mathcal S_0).
\end{equation}
After the band folding, the atomic energy bands at all the momenta in one momentum shell will be folded into the same momentum in the MBZ, allowing us to construct the reducible representation of moir\'e minibands in the 1st MBZ based on atomic energy bands. More concretely, for a certain momentum $\bs k_0 \in A_{\tilde {\bs k}_0}$, we consider the band-folding for the atomic energy bands with the irrep $\Lambda_{\bs k_0}^0$. The bands with the irrep $\Lambda_{\bs k}^0$, which is isomorphic to the irrep $\Lambda_{\bs k_0}^0$, at all $\bs k \in \tilde A_{\bs k_0}$ are folded into the momentum $\tilde {\bs k}_0$ in the 1st MBZ to form an induced representation, 
\begin{equation}
\label{eq: representation after MBZ folding}
   \tilde \Lambda_{\bs k_0} = \Lambda^0_{\bs k_0}\uparrow G_{\tilde{ \bs k}_0},
\end{equation} 
where $\uparrow$ is the induction of representation $\tilde \Lambda_{\bs k_0}$ in the group $G_{\tilde{ \bs k}_0}$ from the representation $\Lambda^0_{\bs k_0}$ of the group $G_{\tilde{ \bs k}_0} \cap G_{\bs k_0}^0 \subset G_{\tilde{ \bs k}_0}$, as defined in TQC \cite{bradlyn2017topological, elcoro2017double, vergniory2017graph}.
$\tilde \Lambda_{\bs k_0}$ is a reducible representation. 
The basis for the induced representation $\tilde \Lambda_{\bs k_0}$ are $\vert \Lambda_{\bs k}^0, n \rangle$ for all $\bs k \in \tilde A_{\bs k_0}$ and $n$, generated by acting all $\mathcal S \in G_{\tilde{ \bs k}_0}$ on the $\vert \Lambda_{\bs k_0}^0, n \rangle$ in the irrep $\Lambda^0_{\bs k_0}$.
As the irreps $\Lambda_{\bs k}^0$ for all $\bs k \in \tilde A_{\bs k_0}$ are transformed to each other by $\mathcal S \in G_{\tilde{\bs k}_0}$, $\tilde \Lambda_{\bs k_0}$ is just $N[\tilde A_{\bs k_0}]$ copies of 
$\Lambda_{\bs k_0}^0$, where $N[\tilde A_{\bs k_0}]$ is the number of different momenta in one momentum shell $\tilde A_{\bs k_0}$. The dimension of $\tilde \Lambda_{\bs k_0}$ is $N[\tilde \Lambda_{\bs k_0}]=N[\tilde A_{\bs k_0}]\times 
N[\Lambda_{\bs k_0}^0]$. The characters of symmetry operators of $G_{ \tilde{\bs k}_0 }$ in $\tilde \Lambda_{\bs k_0}$ are discussed in step 3 below. 

Step 3: we decompose the reducible representation $\tilde \Lambda_{\bs k_0}$ into irreps of the wave vector group $G_{\tilde{\bs k}_0}$ of the moir\'e system using the character table. To decompose $\tilde \Lambda_{\bs k_0}$, we need to evaluate the character of $\tilde \Lambda_{\bs k_0}$, which can be directly connected to the characters of irreps $ \Lambda_{\bs k}^0$
for all $\bs k \in \tilde A_{\bs k_0}$ via Eq.~(\ref{eq: representation after MBZ folding}). 
For a given $\mathcal S \in G_{\tilde {\bs k}_0}$ and a given $\tilde \Lambda_{\bs k_0}$ at the momentum $\bs k_0 \in A_{\tilde{\bs k}_0}$, we can classify all momenta $\bs k \in \tilde A_{\bs k_0}$ into two types: (1) $\mathcal S \in G_{\bs{k}}^0 \cap G_{\tilde {\bs k}_0}$ and (2) $\mathcal S \notin G_{\bs{k}}^0 \cap G_{\tilde {\bs k}_0}$.
For type (1), the representation matrix $\mathcal{D}_{\bs k}^0(\mathcal S )$ and the character $\chi^0_{\bs k}(\mathcal S)$ are defined by Eq. (\ref{eq:repremat_Chara_1}). For type (2), $\bs k$ and $\bs k'=\mathcal S \bs k$ are two different momenta in $G_{\bs{k}}^0$, so all the diagonal components of the representation matrix for $\mathcal S$ must be zero, leading to zero character. With the characters for both types of momenta, we have 
\begin{equation}\label{eq:character_k0_S}
    \tilde{\chi}_{\bs k_0}(\mathcal S) = \sum_{\{\bs k\vert \bs k  \in \tilde{A}_{\bs k_0}, \mathcal S \in G_{\bs{k}}^0 \cap G_{\tilde {\bs k}_0}\}} \chi^0_{\bs k}(\mathcal S).
\end{equation}
Applying Eq.(\ref{eq:character_k0_S}) to all symmetry operators $\mathcal S \in G_{\tilde {\bs k}_0}$, we can construct the characters of all symmetry operators in $G_{\tilde {\bs k}_0}$ for the reducible representation $\tilde \Lambda_{\bs k_0}$, which allows us to decompose $\tilde \Lambda_{\bs k_0}$ through the character table according to the wonderful orthogonality theorem\cite{dresselhaus2007group} into irreps $\Lambda_{\tilde{\bs k}_0, \alpha}$, where $\alpha$ labels different irreps, of $G_{\tilde{\bs k}_0}$, \begin{equation} \label{eq:representation_decomposition}
    \tilde \Lambda_{\bs k_0} = \oplus_{\alpha} c_\alpha \Lambda_{\tilde{\bs k}_0, \alpha}
\end{equation} 
with $c_\alpha$ as the multiple of irreps.

Step 4: we treat the weak moir\'e potential $H_M$ as a perturbation and construct the effective Hamiltonian for the minibands with the irreps in $\tilde \Lambda_{\bs k_0}$. All the states in $\tilde \Lambda_{\bs k_0}$ are degenerate at zero moir\'e potential and the first-order perturbation of the moir\'e potential $H_\text M$ in $\tilde \Lambda_{\bs k_0}$ is 
\begin{equation}
    (H^{(1)}_{\text M,\bs k_0})_{n_1 k_1, n_2 k_2} = \bra{\Lambda_{\bs k_1}^0, n_1} H_\text M \ket {\Lambda_{\bs k_2}^0, n_2}
\end{equation}
with $\bs k_1, \bs k_2 \in \tilde A_{\bs k_0}$. We can diagonalize the effective Hamiltonian $H^{(1)}_{\text M,\bs k_0}$ and the obtained eigen-states are characterized by the irreps $\Lambda_{\tilde{\bs k}_0, \alpha}$ of $G_{\tilde{\bs k}_0}$, which just corresponds to the representation decomposition in Eq.(\ref{eq:representation_decomposition}). 
These eigen-states with different irreps at high symmetry momenta can be ordered in energy and connected along high symmetry lines to construct the band representations of moir\'e minibands by compatibility relations \cite{bradlyn2017topological}.

Step 5: we analyze the topology of the mori\'e minibands using the topological quantum chemistry method\cite{bradlyn2017topological,elcoro2017double,vergniory2017graph} by comparing the band represrentations of moir\'e minibands with the elementary band representation (EBR) from the Bilbao Crystallographic Server \cite{bradlyn2017topological, vergniory2017graph, elcoro2017double}. 

Although we discussed our formalism for the space groups only, time reversal symmetry can also be included in the context of magnetic space groups, and each step in our formalism is still applicable.

\begin{table}
\centering
\includegraphics[width=0.5\columnwidth]{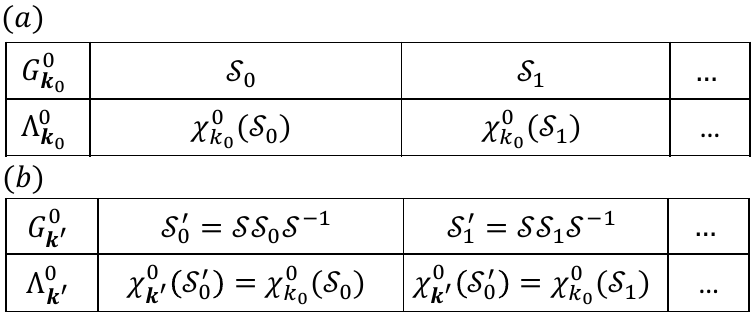}
\caption{
    (a) The characters of symmetry operators $\mathcal S_0, \mathcal S_1, ...$ in the group $G^0_{\bs k_0} \cap G_{ \tilde{\bs k}_0 }$ for the irrep $\Lambda^0_{\bs k_0}$.
    (b) The characters of symmetry operators in the group $G^0_{\bs k'} \cap G_{ \tilde{\bs k}_0}$ for the irrep $\Lambda^0_{\bs k'}$. $\bs k' = \mathcal S \bs k_0$ labels a momentum different from $\bs k_0$, but is connected to $\bs k_0$ by a symmetry operator $\mathcal S \in G_{\tilde k_0}$. 
}
\label{tab: step 2 character table}
\end{table}

\begin{figure}
\centering
\includegraphics[width=\columnwidth]{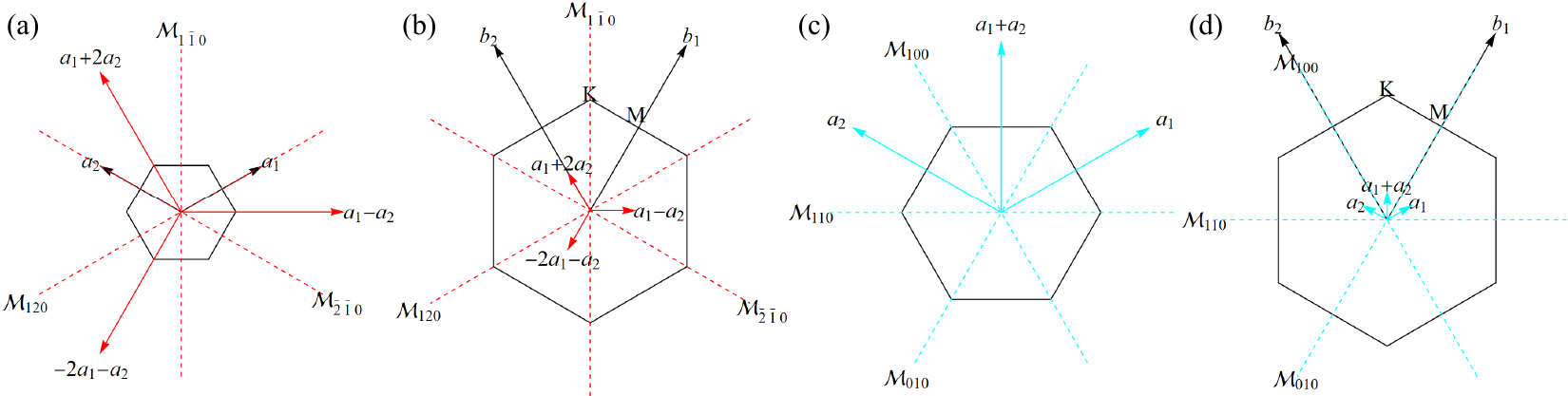}
\caption{
    (a) Mirror symmetries $\mathcal M_{120}, \mathcal M_{\bar 2 \bar 1 0}, \mathcal M_{1\bar 10}$ in the real space. $\bs a_1, \bs a_2$ are the primitive lattice vectors. The red dashed lines are mirror planes of $\mathcal M_{ij0}$ and the red arrows are normal vectors $i\bs a_1 + j\bs a_2$ perpendicular to the mirror plane.
    (b) Mirror symmetries $\mathcal M_{120}, \mathcal M_{\bar 2 \bar 10}, \mathcal M_{1\bar 10}$ in the momentum space space. $\bs b_1, \bs b_2$ are the primitive reciprocal lattice vectors. $i\bs a_1 + j\bs a_2$ indicates the direction of the normal vector perpendicular to the mirror plane of $\mathcal M_{ij0}$.
    (c)(d) Mirror symmetries $\mathcal M_{110}, \mathcal M_{100}, \mathcal M_{010}$ in the real (momentum) space.
    The blue dashed lines are mirror planes of $\mathcal M_{ij0}$ with the normal vector in the direction of blue arrows $i \bs a_1 + j \bs a_2$.
}
\label{fig: moire mirror}
\end{figure}

\section{Moir\'e Rashba systems}
\label{sec:MoireRashba}
\subsection{Moir\'e TQC analysis}\label{sec: moire rashba tqc}
In this section, we will apply our general formalism developed in Sec.\ref{sec:moire TQC} to the Rashba moir\'e system as an example to show that the band inversion between the states derived from the 1st and 2nd MBZ leads to the exchange of the states with different symmetry irreps at $\Gamma$ and $K$, giving rise to nontrivial band topology.
The results of each step in Sec.\ref{sec:moire TQC} is shown in \figref{fig: moire TQC}, which will be described in details below. 

Step 1: we first identify the momentum shells folded to $\Gamma$ and $K$ from the 1st and 2nd extended MBZ. The space group for the Rashba model is $G^0 = C_{\infty v} \otimes G^\text T$, where $G^\text T$ is the continuous translation group and $C_{\infty v}$ is the full rotation group about z-axis together with in-plane mirrors.
The space group $G^\text M$ of the moir\'e potential $H_\text M$ is chosen to be $P6/mmm$, corresponding to the point group $D_{6h}$\cite{de2021layer}. 
The space group for the whole system is $P6mm$ from $G = G^\text M \cap G^0$ corresponding to the point group $C_{6v}$.

We first consider the momentum set $A_{\Gamma}$, namely the momenta $\bs k_0$ in ABZ that can be folded to $\tilde{\bs k}_0 = \Gamma$ in MBZ.
The wave vector group at $\Gamma$ for $G$ is $G_{\Gamma} = C_{6v}$ with the symmetry operators, including the six-fold, three-fold, two-fold rotation symmetries about the z axis $\mathcal C_{6z}, \mathcal C_{3z}, \mathcal C_{2z}$, respectively, and the mirror symmetries $\mathcal M_{ij0}$ with the normal vector along the direction $i \bs a_1^\text M + j \bs a_2^\text M$ shown in the real space and momentum space in Fig.~\ref{fig: moire mirror}, where $\bs a_1^\text M = L_\text M (\sqrt 3 / 2 , 1/2), \bs a_2^\text M = L_\text M (-\sqrt 3 / 2 , 1/2)$ are primitive vectors for the moir\'e unit cell. Here we follow the notation in the Bilbao Crystallographic Server \cite{aroyo2014brillouin, tasci2012introduction}. 
Tab.~\ref{tab: irreps at G}(a) gives the irreps only appearing in the double group of $C_{6v}$. In Tab.~\ref{tab: irreps at G}(a), symmetry operators belonging to one conjugacy class are put into the same column.
From the definition of \eqnref{eq:shell of momenta}, the first two momentum shells are given by  
\begin{equation} \label{eq:momentum_shell_Gamma}
    \begin{split}
        \tilde A_{\Gamma} &= \{\Gamma\} \\ 
        \tilde A_{\Gamma^1} &= \{\pm \bs b_1^\text M, \pm \bs b_2^\text M, \pm(\bs b_1^\text M -\bs b_2^\text M)\},
    \end{split}
\end{equation}
where the primitive reciprocal lattice vectors are given by $\bs b_1^\text M = \frac{4 \pi} {\sqrt 3 L_\text M} (\sqrt 3/2, 1/2), \bs b_2^\text M = \frac{4 \pi}{\sqrt 3 L_\text M} (-\sqrt 3/2, 1/2)$.
The momenta in $\tilde A_{\Gamma_1}$ are obtained from $\Gamma^1=\bs b_1^\text M$ by
\begin{equation}\label{eq: momentum connection to G1}
    \bs b_2^\text M = \mathcal C_{6z} \Gamma^1  \quad \bs b_2^\text M - \bs b_1^\text M = \mathcal C_{3z} \Gamma^1 \quad - \bs b_1^\text M = \mathcal C_{2z} \Gamma^1 \quad - \bs b_2^\text M = \mathcal C_{3z}^{-1} \Gamma^1 \quad \bs b_1^\text M - \bs b_2^\text M = \mathcal C_{6z}^{-1} \Gamma^1.
\end{equation}
Next, we consider the momentum shells that are connected to $K$.
The wave vector group at $K$ is $G_{K} = C_{3v}$ with the symmetry operators and double group irreps listed in Tab.~\ref{tab: irreps at K}(a).
Here we choose $K = \bs b_1^\text M / 3+ \bs b_2^\text M /3 $ in the 1st MBZ and $K^1 = 4 \bs b_1^\text M /3 - 2 \bs b_1^\text M /3 $ in the 2nd MBZ to construct the first two momentum shells, following step 1 of \figref{fig: moire TQC}. Similarly from \eqnref{eq:shell of momenta}, the first two momentum shells are 
\begin{equation} \label{eq:momentumShell_K}
    \begin{split}
        \tilde A_{K} &= \{ \bs b_1^\text M / 3+ \bs b_2^\text M /3,  -2 \bs b_1^\text M / 3+ \bs b_2^\text M /3, \bs b_1^\text M / 3- 2 \bs b_2^\text M /3\} \\ 
        \tilde A_{K^1} &= \{ 4 \bs b_1^\text M /3 - \bs b_2^\text M /3, -2 \bs b_1^\text M /3 + 4 \bs b_2^\text M /3, - 2 \bs b_1^\text M /3 - 2\bs b_2^\text M/3 \}.
    \end{split}
\end{equation}

\begin{table}
\centering
\includegraphics[width=\columnwidth]{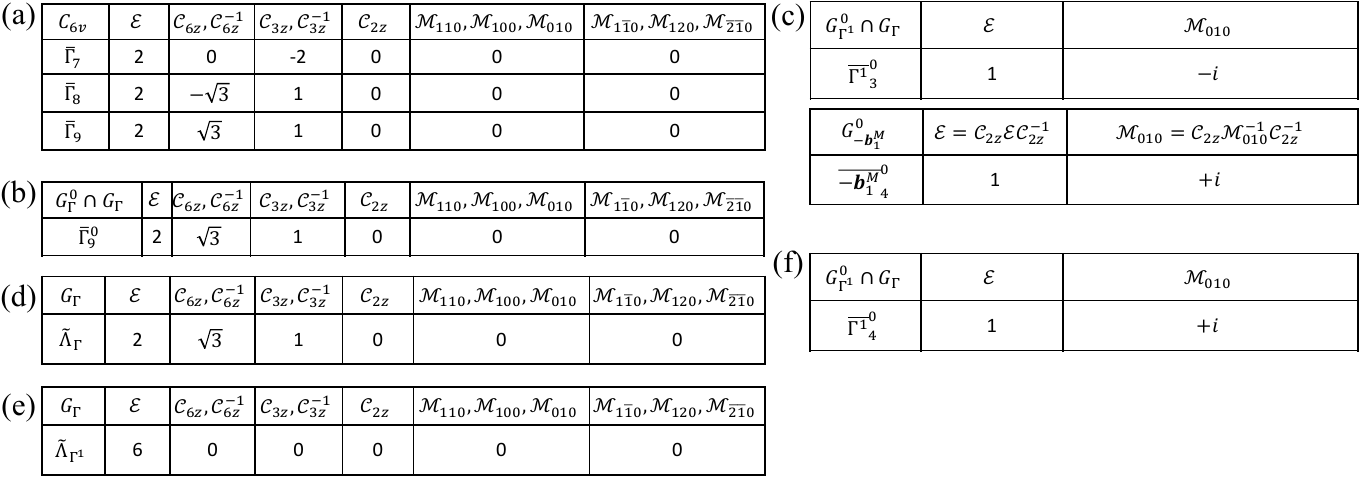}
\caption{
    (a) The character table for the wave vector group $G_\Gamma = C_{6v}$.
    (b) The characters of symmetry operators in the group $G_\Gamma^0 \cap G_\Gamma = C_{6v}$ for the irrep $\Lambda^0_{\Gamma}$ of the unfolded Rashba model at $\Gamma$. 
    (c) The characters of the symmetry operators in the group $G_{\Gamma^1}^0 \cap G_\Gamma = C_m (G_{-\bs b_1^\text M}^0 \cap G_\Gamma = C_m)$ for the irrep $\Lambda^0_{\Gamma^1} (\Lambda^0_{-\bs b_1^\text M})$ of the unfolded Rashba model at $\Gamma^1 (-\bs b_1^\text M)$.
    $\Gamma^1 = \bs b_1^\text M, -\bs b_1^\text M$ are two momenta invariant under $\mathcal M_{010}$.
    (d)(e) The characters of the symmetry operators in the group $G_{\Gamma}$ for the folded representations $\tilde \Lambda_\Gamma, \tilde \Lambda_{\Gamma^1}$.
    (f) The characters of the symmetry operators in the group $G_{\Gamma^1}^0 \cap G_\Gamma$ for the irrep $\bar{\Gamma^1}^0_4$ of the unfolded Rashba model at $\Gamma^1$.
}
\label{tab: irreps at G}
\end{table} 

\begin{table}
\centering
\includegraphics[width=0.8\columnwidth]{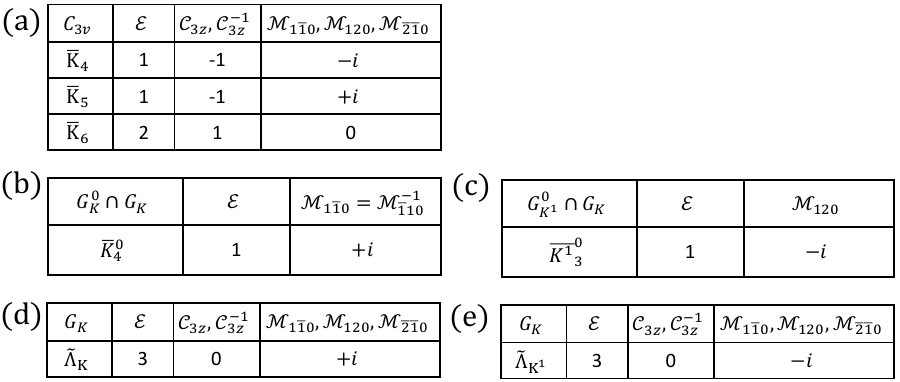}
\caption{
    (a) The character table for the wave vector group $G_K = C_{3v}$.
    (b)(c) The characters of symmetry operators in the group $G_K^0 \cap G_K = C_m (G_{K^1}^0 \cap G_k = C_m)$ for the irrep $\Lambda^0_{K} = \bar K_4^0 (\Lambda^0_{K^1} =  \bar {K^1}_3^0 )$ of the unfolded Rashba model at $K$.
    (d)(e) The characters of symmetry operators in the group $G_K$ for folded representations $\tilde \Lambda_K(\tilde \Lambda_{K^1})$.
}
\label{tab: irreps at K}
\end{table} 


Step 2: we will derive the representations of states folded from $\tilde A_{\Gamma}$, $\tilde A_{K}$ in the 1st MBZ and $\tilde A_{\Gamma^1}$, $\tilde A_{K^1}$ in the 2nd MBZ. At $\Gamma$, the wave vector groups are $G^0_{\Gamma} = C_{\infty v}$ of the unfolded Rashba model and $G_{\Gamma} = C_{6v}$ for the moir\'e systems. Thus, we choose the group $G^0_{\Gamma} \cap G_{\Gamma} = C_{6v}$ to label atomic bands at $\Gamma$.
The symmetry irrep of the unfolded Rashba model under $G^0_{\Gamma} \cap G_{\Gamma}$ is $\Lambda^0_{\Gamma} = \bar{\Gamma}_{9}^0$. 
Here we emphasize that we use the notation $\bar{\Gamma}$ (with a bar on top of the high symmetry momentum) to label the irreducible representation, which is different from $\Gamma$ that labels the momentum, and we follow this convention below. 
The characters of symmetry operators in \addKJ{$\bar{\Gamma}^0_{9}$} are listed in Tab.~\ref{tab: irreps at G}(b).
Since there is only $\Gamma$ in $\tilde A_{\Gamma}$, the representations is $\tilde \Lambda_\Gamma = \bar{\Gamma}^0_{9}$ in MBZ.
At $\Gamma^1$, the unfolded Rashba model has the wave vector group $G^0_{\Gamma^1} = C_{m}$ with only in-plane mirror symmetry $\mathcal M_{010}$.
The irrep of the lower energy band of the unfolded Rashba model at $\Gamma^1$ is $\bar{\Gamma^1}^0_3$ with the characters $\chi^0_{\Gamma^1}(\mathcal S_0)$ for the symmetry operators $\mathcal S_0$ in the group $G^0_{\Gamma^1} \cap G_{\Gamma} = C_m$ listed in Tab.~\ref{tab: irreps at G}(c).
Along the line $\Gamma-\Gamma^1$, $\bar{\Gamma}_{9}^0$ splits by $ \bar{\Gamma}_{9}^0 = \bar{\Gamma^1}^0_3 \oplus \bar{\Gamma^1}^0_4$
, where $\bar{\Gamma^1}^0_3$ and $\bar{\Gamma^1}^0_4$ are two double group irreps at $\Gamma^1$ with opposite mirror eigenvalues $-i$ and $+i$, respectively, under the mirror operator $\mathcal M_{010}$ with the characters listed in Tab.~\ref{tab: irreps at G}(c)(f).
The Rashba Hamiltonian is $H_\text R(\bs  k)  = ak^2 \sigma_0 -\lambda (\bs k \times \bs \sigma)\cdot \hat z$ as \eqnref{eq: rashba ham} in the main text and the mirror symmetry operator with the mirror plane normal vector $\hat z \times \hat k$ is $\mathcal M_{\hat z \times \hat k} = - i (\hat z \times \hat k)\cdot \bs \sigma$, where $\hat z, \hat k$ are unit vectors along the $z,\bs k$ direction, respectively.
For the Rashba strength $\lambda > 0$, we find $\bar{\Gamma^1}^0_3$ with the mirror eigenvalue $-i$ has the lower energy $a \vert \bs b_1^\text M\vert^2 -\lambda \vert \bs b_1^\text M\vert$.
Next, the characters at $\bs k'\in \tilde A_{\Gamma^1}$ of $\Lambda^0_{\bs k'}$ are derived in the Tab.~\ref{tab: irreps at G}(c) following Tab.~\ref{tab: step 2 character table}.
From \eqnref{eq: representation after MBZ folding}, we obtain $\tilde \Lambda_{\Gamma^1} = \Lambda_{\Gamma^1}^0 \uparrow G_\Gamma$.

Similarly, at $K, K^1$, the unfolded Rashba model has the wave vector group $G_{K} \cap G_{K}^0=G_{K} \cap G_{K^1}^0= C_{m}$
and the corresponding representations $\tilde \Lambda_{K}, \tilde \Lambda_{K^1}$ can be constructed from the irreps $\bar{\bs K}^0_4, \bar{\bs K^1}^0_3$ at $\tilde A_{K}, \tilde A_{K^1}$ with their characters given in Tab.~\ref{tab: irreps at K}. 


Step 3: we decompose the representation $ \tilde \Lambda_\Gamma, \tilde \Lambda_{\Gamma^1}$ into irreps of the wave vector group $G_\Gamma$ at $\Gamma$ and $ \tilde \Lambda_K, \tilde \Lambda_{K^1}$ into irreps of the wave vector group $G_K$ at $K$ using the character table.
We evaluates the characters for the representations $\tilde \Lambda_\Gamma$ and $\tilde \Lambda_{\Gamma^1}$ in Tab.~\ref{tab: irreps at G}(d) and (e), and then decompose them into irreps according to the character table of $G_{\bs \Gamma} = C_{6v}$ in Tab.~\ref{tab: irreps at G}(a).
For $\tilde \Lambda_\Gamma$, $\tilde \chi_\Gamma(\mathcal S) = \chi_\Gamma^0(\mathcal S)$ for $\mathcal S\in G_\Gamma = C_{6v}$
using \eqnref{eq:character_k0_S}, as shown in Tab.~\ref{tab: irreps at G}(d).
By comparing Tab.~\ref{tab: irreps at G}(d) with the character table in Tab.~\ref{tab: irreps at G}(a), the wonderful orthogonality theorem gives
\begin{equation}
    \tilde \Lambda_\Gamma = \bar \Gamma_9.
\end{equation}

For $\tilde \Lambda_{\Gamma^1}$, the characters of all symmetry operators in the wave vector group $G_{\Gamma}$ are summarized in Tab.~\ref{tab: irreps at G}(e). Below we take the characters $\tilde \chi_{\Gamma^1}(\mathcal E)$ and $\tilde \chi_{\Gamma^1}(\mathcal M_{010})$ as examples.
The identity operator $\mathcal E$ appears in $G^0_{\bs k} \cap G_{\Gamma}$ for any $\bs k \in \tilde A_{\Gamma^1}$, and thus $\tilde \chi_{\Gamma^1}(\mathcal E) = 6$ from \eqnref{eq:character_k0_S}, as there are six momenta in $\tilde A_{\Gamma^1}$.
For the symmetry operator $\mathcal M_{010}$, only the momenta $\Gamma^1=\bs b_1^\text M$ and $-\bs b_1^\text M$ are invariant, while the other momenta in $\tilde A_{\Gamma^1}$ (Eq.~\ref{eq:momentum_shell_Gamma}) are not invariant. 
The mirror eigen-values for the irrep $\bar{\Gamma^1}^0_3$ and $\bar{-\bs b_1^\text  M}^0_4$ are $-i$ and $+i$, respectively. Thus, we conclude $\tilde \chi_{\Gamma^1}(\mathcal M_{010}) = 0$ from \eqnref{eq:character_k0_S}. 
$\mathcal M_{100}, \mathcal M_{110}, \mathcal M_{100}^{-1}, \mathcal M_{110}^{-1}$ belong to the same conjugacy class in $G_\Gamma$ and appear in the same column in Tab.~\ref{tab: irreps at G}(c). 
The characters for other symmetry operators in $G_\Gamma$ can be constructed in a similar manner. By comparing Tab.~\ref{tab: irreps at G}(e) and Tab.~\ref{tab: irreps at G}(a), the wonderful orthogonality theorem gives
\begin{equation} \label{eq:rep_decomp_Gamma1}
    \tilde \Lambda_{\Gamma_1} = \bar \Gamma_7 \oplus \bar \Gamma_8 \oplus \bar \Gamma_9.
\end{equation}
Next, $\tilde \Lambda_K, \tilde \Lambda_{K^1}$ is decomposed with the character table in Tab.~\ref{tab: irreps at K}(d)(e) derived from Tab.~\ref{tab: irreps at K}(b)(c) following   \eqnref{eq:character_k0_S}.
By comparing Tab.~\ref{tab: irreps at K}(d)(e) and Tab.~\ref{tab: irreps at K}(a), we obtain
\begin{equation}
    \tilde \Lambda_{K} = \bar K_5 \oplus \bar K_6 , \quad \quad \tilde \Lambda_{K^1} = \bar K_4 \oplus \bar K_6.
\end{equation}

Step 4: we write down effective Hamiltonian for the mini-bands of each irrep obtained from Step 3 at high symmetry momenta by treating the moir\'e potential with the perturbation theory, solve the eigen-energy problem for the effective Hamiltonian of each irrep, arrange them in energy and then connect them into energy bands.
Starting from the unperturbed Rashba model in \eqnref{eq: rashba ham} of the main text, the energies and eigenstates are
\begin{equation}\label{eq:Rashba_eigen1}
    \vert \bs k, \pm \rangle = 
    \frac{1}{\sqrt 2}
    \begin{pmatrix}
        1 \\
        \mp \frac{i k_x - k_y}{k}
    \end{pmatrix}, \quad
    E_\text R(\bs k, \pm)= ak^2 \pm \lambda k.
\end{equation}

As the first momentum shell $\tilde A_\Gamma$ only has one momentum, $H_\text M$ does not split the degeneracy of  $\tilde \Lambda_\Gamma$ protected by the time reversal $\mathcal T$. Thus, the irrep $\tilde \Lambda_\Gamma$ has the energy 
\begin{equation}
    E_{\Gamma}(\bar \Gamma_9) = 0
\end{equation}
for both spin states.

In contrast, $H_\text M$ can split the degenerate states folded from the 2nd momentum shell $\tilde A_{\Gamma^1}$. We consider the lower eigen-energy state $|\bs k,-\rangle$ of $H_R$ in Eq.(\ref{eq:Rashba_eigen1}), and on the basis $\vert \bs b_1^\text M,-\rangle, \vert \bs b_2^\text M,-\rangle, \vert \bs b_2^\text M - \bs b_1^\text M,-\rangle, \vert -\bs b_1^\text M,-\rangle, \vert -\bs b_2^\text M,-\rangle, \vert \bs b_1^\text M-\bs b_2^\text M,-\rangle$, we can evaluate the matrix element $\langle \bs k_1, - \vert H_\text R + H_\text M \vert \bs k_2, - \rangle$ 
and construct effective Hamiltonian for the irrep $\tilde \Lambda_{\Gamma^1}$ as
\begin{equation} \label{eq:gamma rashba}
\begin{split}
    H_{\Gamma^1}  = \begin{pmatrix}
        E_\text R(\bs b_1^\text M,-) & \Delta_1 e^{i\pi/6}\sqrt 3/2 & 0 & 0 & 0 & \Delta_1 e^{-i\pi/6}\sqrt 3/2\\
        \Delta_1 e^{-i\pi/6}\sqrt 3/2& E_\text R(\bs b_2^\text M,-) & \Delta_1 e^{i\pi/6}\sqrt 3/2 & 0 & 0 & 0\\
        0 & \Delta_1 e^{-i\pi/6}\sqrt 3/2 & E_\text R(\bs b_2^\text M - \bs b_1^\text M,-) & \Delta_1 e^{i\pi/6}\sqrt 3/2 & 0 & 0\\
        0 & 0 &\Delta_1 e^{-i\pi/6}\sqrt 3/2 & E_\text R(-\bs b_1^\text M,-) & \Delta_1 e^{i\pi/6}\sqrt 3/2  & 0\\
        0 & 0 & 0 &\Delta_1 e^{-i\pi/6}\sqrt 3/2 & E_\text R(-\bs b_2^\text M,-) & \Delta_1 e^{i\pi/6}\sqrt 3/2 \\
        \Delta_1 e^{i\pi/6}\sqrt 3/2 & 0 & 0 & 0 &\Delta_1 e^{-i\pi/6}\sqrt 3/2 & E_\text R(\bs b_1^\text M-\bs b_2^\text M,-) 
    \end{pmatrix},
\end{split}
\end{equation}
where $\bs k_1, \bs k_2 \in \tilde A_{\Gamma^1}$ and $E_\text R(\Gamma^1, -)=a \vert \bs b_1^\text M \vert^2 - \lambda \vert \bs b_1^\text M \vert$. The eigen-energies of $H_{\Gamma^1}$ are
\begin{equation}
    E_{\Gamma^1}(\bar \Gamma_8) = E_\text R(\Gamma^1, -) - 3\Delta_1 / 2 \quad
    E_{\Gamma^1}(\bar \Gamma_7) = E_\text R(\Gamma^1, -) \quad
    E_{\Gamma^1}(\bar \Gamma_9) = E_\text R(\Gamma^1, -) + 3\Delta_1 / 2
\end{equation}
with double degeneracy as required by time reversal. 
Here $\bar \Gamma_{7,8,9}$ in the bracket after $E_{\Gamma^1}$ gives the irrep for each eigen-state, which can be determined from the eigen-values of $\mathcal C_{6z}$ with the transformation matrix
\begin{equation}
    \mathcal C_{6z}(\tilde \Lambda_{\Gamma^1}) = 
    \begin{pmatrix}
        0 & 0 & 0 & 0 & 0 & e^{-i\pi / 6} \\
        e^{-i\pi / 6} & 0 & 0 & 0 & 0 & 0 \\
        0 & e^{-i\pi / 6} & 0 & 0 & 0 & 0 \\
        0 & 0 & e^{-i\pi / 6} & 0 & 0 & 0 \\
        0 & 0 & 0 & e^{-i\pi / 6} & 0 & 0 \\
        0 & 0 & 0 & 0 & e^{-i\pi / 6} & 0 
    \end{pmatrix}.
\end{equation}
The eigen-energy splittings of the $\tilde \Lambda_{\Gamma^1}$ states from the above Hamiltonian $H_{\Gamma^1}$ are consistent with the representation decomposition in Eq.(\ref{eq:rep_decomp_Gamma1}). 
For $\Delta_1>0$, the minibands with the irrep $\bar \Gamma_8$ have the lowest energy for $\tilde \Lambda_{\Gamma^1}$, which is consistent with the Fig. 2f in the main text.

Next, we determine the energy order of minibands with different irreps in $\tilde \Lambda_{K}, \tilde \Lambda_{K^1}$.
At $K$, the first order perturbation Hamiltonian of $H_\text M$ in the basis of states at $\tilde A_K$ is
\begin{equation}
    H_K = 
    \begin{pmatrix}
        E_\text R(K, -) & \Delta_1 e^{i \pi /3} /2 & \Delta_1 e^{-i \pi /3} /2 \\
        \Delta_1 e^{- i \pi /3} /2 & E_\text R(-\bs b_1^\text M 2/3 + \bs b_1^\text M/3, -) & \Delta_1 e^{i \pi /3} /2 \\
        \Delta_1 e^{i \pi /3} /2 & \Delta_1 e^{- i \pi /3} /2 & E_\text R(\bs b_1^\text M /3 - \bs b_1^\text M2/3, -) 
    \end{pmatrix},
\end{equation}
where $E_\text R(K, -)= a \vert \bs b_1^\text M/3 + \bs b_2^\text M/3 \vert^2 - \lambda \vert \bs b_1^\text M/3 + \bs b_2^\text M/3 \vert $.
The energies of $H_K$ are
\begin{equation}
    E_K(\bar K_5) = E_\text R(K, -) - \Delta_1 \quad E_K(\bar K_6) = E_\text R(K, -) + \Delta_1. 
\end{equation}
Here the $E_K(\bar K_5)$ state has no degeneracy while the $E_K(\bar K_6)$ states are doubly degenerate. For $\Delta_1>0$, $E_K(\bar K_5) < E_K(\bar K_6)$.

At $K^1$, the first order perturbation Hamiltonian of $H_\text M$ in the basis of states at $\tilde A_{K^1}$ is zero because we made the first momentum shell approximation for $H_\text M$, so that $\bs k_2 - \bs k_1$ is not within the $\bs g$ summation in $H_\text M$ (Eq.\ref{eq: moire potential main}) of the main text for any $\bs k_1, \bs k_2 \in \tilde A_{K^1}$. Thus, we need to consider the second order perturbation to get the effective Hamiltonian, 
\begin{equation}
\begin{split}
    & (H_{K^1})_{\bs k_1, \bs k_2}  = \langle \bs k_1, - \vert H_\text R \vert \bs k_2, - \rangle + \sum_{\bs g \in \{ \pm \bs b_1^\text M, \pm \bs b_2^\text M, \pm (\bs b_1^\text M - \bs b_2^\text M )\}, n=\pm} \frac{ \langle \bs k_1, - \vert H_\text M \vert \bs k_1 + \bs g, n \rangle \langle \bs k_1 + \bs g, n \vert H_\text M \vert \bs k_2, - \rangle }{E_\text R(K^1, -) - E_\text R(\bs k_1 + \bs g, n)},  \\
    & H_{K^1} = \begin{pmatrix}
        E_\text R(K^1,-) + \Delta_1^2 f_{d}^{(2)}(K^1) & \Delta_1^2 e^{i\pi/3}f_{od}^{(2)}(K^1)  & \Delta_1^2 e^{-i\pi/3}f_{od}^{(2)}(K^1)  \\
        \Delta_1^2 e^{-i\pi/3}f_{od}^{(2)}(K^1) & E_\text R(-\bs b_1^\text M 2/3 - \bs b_2^\text M 4/ 3,-) + \Delta_1^2 f_{d}^{(2)}(K^1) & \Delta_1^2 e^{i\pi/3}f_{od}^{(2)}(K^1) \\
        \Delta_1^2 e^{i\pi/3}f_{od}^{(2)}(K^1) & \Delta_1^2 e^{-i\pi/3}f_{od}^{(2)}(K^1) & E_\text R(-\bs b_1^\text M 2/3 - \bs b_2^\text M 2/ 3,-) +\Delta_1^2 f_{d}^{(2)}(K^1) 
    \end{pmatrix},
\end{split}
\end{equation}
where $\bs k_1, \bs k_2 \in \tilde A_{K^1}$,  $E_\text R(K^1,-)= a \vert 4\bs b_1^\text M/3 - \bs b_2^\text M/3 \vert^2 - \lambda \vert 4\bs b_1^\text M/3 - \bs b_2^\text M/3 \vert$, and 
\begin{equation}\label{eq:pert_2ndorder_K1}
\begin{split}
    f_{od}^{(2)}(K^1) &= \frac{3}{4}\frac{1}{E_\text R(K^1,-) - E_\text R(K,-)} - \frac{1}{4}\frac{1}{E_\text R(K^1,-) - E_\text R(K,+)} \\
    f_{d}^{(2)}(K^1) &= \sum_{\bs g \in \{ \pm \bs b_1^\text M, \pm \bs b_2^\text M, \pm (\bs b_1^\text M - \bs b_2^\text M )\}, n=\pm} \frac{ \langle  K^1, - \vert K^1 + \bs g, n \rangle \langle K^1 + \bs g, n \vert K^1, - \rangle }{E_\text R(K^1, -) - E_\text R(K^1 + \bs g, n)}.
\end{split}
\end{equation}
The energies of $H_{K^1}$ are
\begin{equation}
    E_{K^1}(\bar K_4) = E_\text R(K^1, -)  + \Delta_1^2 f_{d}^{(2)}(K^1) - 2 \Delta_1^2f_{od}^{(2)}(K^1)  \quad E_K(\bar K_6) = E_\text R(K^1, -)  + \Delta_1^2 f_{d}^{(2)}(K^1) + \Delta_1^2 f_{od}^{(2)}(K^1).
\end{equation}
In the weak moir\'e potential limit $\Delta_1\rightarrow 0$, when the $\tilde \Lambda_{K^1}$ states from the 2nd momentum shell with the dominant energy $E_\text R(K^1,-)$ and the $\tilde \Lambda_{K}$ states from the 1st momentum shell with $E_\text R(K,-)$ exchange in energy, $f_{od}^{(2)}(K^1)$ has a sign change as the energy difference $E_\text R(K^1,-)-E_\text R(K,-)$ appears in the denominator of Eq.(\ref{eq:pert_2ndorder_K1}), so that $E_{K^1}(\bar K_4)$ and $E_{K^1}(\bar K_6)$ in $\tilde \Lambda_{K^1}$ also exchange their energy orders.
For $E_\text R(K^1,-) < E_\text R(K,-) < E_\text R(K,+)$, $f_{od}^{(2)}(K^1)<0$ and $E_{K^1}(\bar K_4) > E_{K^1}(\bar K_6)$. This corresponds to the blue region after the phase transition depicted by the black line at $L_M=4 L_R$ in Fig.~\ref{fig: Rashba topological phase diagram}a of main text. 

\begin{table}
\centering
\includegraphics[width=0.4\columnwidth]{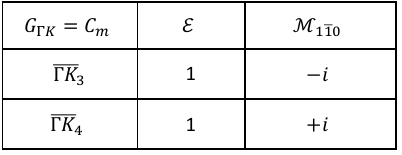}
\caption{
    The character table for the wave vector group $G_{\Gamma K} = C_{m}$.
}
\label{tab: irreps at GK}
\end{table} 

\begin{figure}
\centering
\includegraphics[width=\columnwidth]{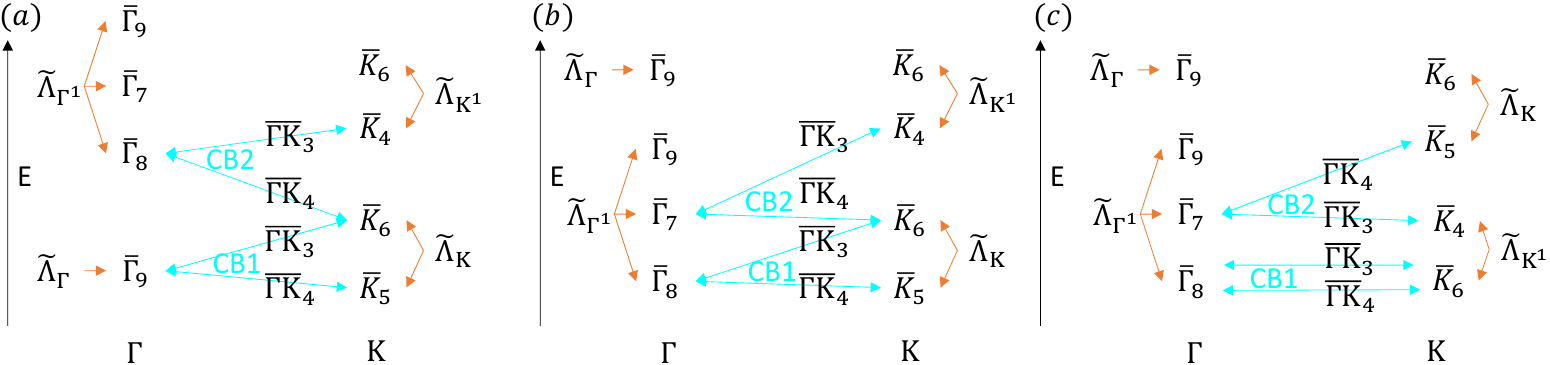}
\caption{
    (a)(b)(c) The irreps at $\Gamma,K$ for the lowest two conduction minibands CB1 and CB2 determined by the compatibility relation for the parameter $1/\sqrt 3 < \sqrt 3 L_\text M/4L_\text R < 1, 1 < \sqrt 3 L_\text M/4L_\text R <\sqrt 3, \sqrt 3< \sqrt 3 L_\text M/4L_\text R$, respectively. The orange lines are the split of the folded representations by a weak moir\'e potential. The blue lines connect the irreps at $\Gamma, K$ through the same irrep at the $\Gamma-K$ line, namely the compatibility relation.
}
\label{fig: compatibility relation}
\end{figure}

Next, different irrep states at high symmetry momenta are connected into bands by the compatibility relation.
According to the compatibility relation along the $\Gamma-K$ line, all double group irreps $\bar \Gamma_7, \bar \Gamma_8, \bar \Gamma_9$ at $\Gamma$ are split into $\overline{\Gamma K}_3\oplus \overline{\Gamma K}_4$ and $\bar K_4, \bar K_5, \bar K_6$ becomes $\overline{\Gamma K}_3, \overline{\Gamma K}_4, \overline{\Gamma K}_3\oplus \overline{\Gamma K}_4$, respectively. Here we use the notation $\overline{\Gamma K}$ to label the irreps along the $\Gamma$-$K$ lines.
The character table for the wave vector group of momenta at $\Gamma-K$ is listed in Tab.~\ref{tab: irreps at GK}.
The irreps at $\Gamma, K$ for different Rashba SOC strength are connected as shown in \figref{fig: compatibility relation}. Here we focus on the band connection of the lowest conduction bands, denoted as CB1, and the second lowest conduction bands, denoted as CB2, based on the above discussion on compatibility relation. 
For a small Rashba SOC parameter $\lambda \ll 1$ (or equivalently $L_M/L_R \ll 1$), the irrep at $\Gamma$ for the lowest energy miniband CB1 comes from $\tilde \Lambda_{\Gamma}$ that belong to the $\bar \Gamma_9$ irrep and
the irrep at $K$ comes from $\tilde \Lambda_{K}$ with two irreps $\bar K_5$ and $\bar K_6$. For $\Delta_1>0$, $E_K(\bar K_5) < E_K(\bar K_6)$ so that the lowest energy bands at $K$ belong to the 1D $\bar K_5$ irrep. The 2D irrep $\bar \Gamma_9$ from $\tilde \Lambda_{\Gamma}$ is connected to 1D $\bar K_5$ for one state while the other state in $\bar \Gamma_9$ is connected to one state in 2D $\bar K_6$ in $\tilde \Lambda_{K}$, forming the band representations of CB1. Thus, CB1 must have band touching with higher energy minibands, which is shown as the semimetal phase in the phase diagram for Fig.2a of the main text. 
When the Rashba parameter $\lambda$ is increased so that $1/\sqrt 3 < \sqrt 3 L_\text M/4L_\text R < 1$, 
$E_\text R(K, +) > E_\text R(K^1, -)$, and the irreps $\bar \Gamma_9$ from $\tilde \Lambda_{\Gamma}$ and $\bar \Gamma_8$ from $\tilde \Lambda_{\Gamma^1}$ are connected to $\bar K_5$ and $\bar K_6$ from $\tilde \Lambda_{K}$ and $\bar K_4$ from $\tilde \Lambda_{K^1}$ as shown in \figref{fig: compatibility relation}(a), so that the CB1 still have band touching with CB2 and thus remains in the semi-metal phase. However, the band character of CB2 is changed due to the band inversion between the $E_\text R(K, +)$ bands and $E_\text R(K^1, -)$ bands, as compared to the case of $L_M/L_R \ll 1$. 
The miniband inversion from different momentum shells first occurs at $\Gamma$ between the $\tilde \Lambda_\Gamma$ state from the 1st MBZ and the $\tilde \Lambda_{\Gamma^1}$ states from the 2nd MBZ for $\sqrt 3 L_\text M/4L_\text R = 1$. As shown in \figref{fig: compatibility relation}(b), when $1 < \sqrt 3 L_\text M/4L_\text R < \sqrt{3}$, the $\bar \Gamma_8$ and $\bar \Gamma_7$ states from $\tilde \Lambda_{\Gamma^1}$ have lower energy than the $\tilde \Lambda_{\Gamma}$ states, so they are connected to $\bar K_5, \bar K_6$ from $\tilde \Lambda_{K}$ and $\bar K_4$ from $\tilde \Lambda_{K^1}$, forming the band representations of CB1 and CB2, which are still connected via band touching. With further increasing the Rashba SOC $\lambda$, the miniband inversion from different momentum shells occurs at $K$, between the $\tilde \Lambda_{K}$ states from the 1st momentum shell and the $\tilde \Lambda_{K^1}$ states from the 2nd momentum shell for $\sqrt 3 L_\text M/4L_\text R = \sqrt{3}$. As shown in \figref{fig: compatibility relation}(c), when $\sqrt 3 L_\text M/4L_\text R > \sqrt{3}$, the $\tilde \Lambda_{K^1}$ states have the lower energy than the $\tilde \Lambda_{K}$ states and the 2D irrep $\bar K_6$ states have the lowest energy. The 2D irrep $\bar \Gamma_8$ from $\tilde \Lambda_{\Gamma^1}$ is connected to the 2D $\bar K_6$ from $\tilde \Lambda_{K^1}$, forming the band representations of CB1, which are isolated from other minibands, as shown in \figref{fig: compatibility relation}(c). The band representation of CB1 corresponds to the blue region of \figref{fig: Rashba topological phase diagram}(a) of the main text, in which the irrep at $\Gamma$ for the lowest energy miniband CB1 comes from $\tilde \Lambda_{\Gamma^1}$ and the irrep at $K$ comes from $\tilde \Lambda_{K^1}$.

Step 5: we compare the band representations at high symmetry points shown in \figref{fig: compatibility relation} with EBR to determine its topology.
For \figref{fig: compatibility relation}(a), as the $\bar \Gamma_9$ irrep for the lowest energy states at $\Gamma$ is 2D while the $\bar K_5$ irrep at $K$ is 1D, CB1 has to be connected to CB2 (the second lowest energy band) at $K$, thus forming a semi-metal phase. CB1 and CB2 together from a band representation that is the same as the EBR $\bar E_1@2b$ listed in the table of the step 5 in Fig.\ref{fig: moire TQC}. Therefore, we expect the CB1 and CB2 together to be topologically trivial. In \figref{fig: compatibility relation}(b) after the band inversion between $\tilde \Lambda_{\Gamma}$ and $\tilde \Lambda_{\Gamma^1}$ at $\Gamma$, CB1 is still in the semi-metal phase. CB1 and CB2 together do not correspond to any EBR and thus become topological after band inversion. In this sense, it is a topological phase transition due to the irrep exchange between the 1st MBZ and 2nd MBZ at $\Gamma$.
For \figref{fig: compatibility relation}(c) after the band inversion between $\tilde \Lambda_{K}$ and $\tilde \Lambda_{K}^1$ at $K$, CB1 becomes isolated and has the same irreps as EBR $\bar E_2 @ 1a$. However, in this case CB1 is still topologically nontrivial, which can be characterized by nontrivial $\mathbb Z_2$ number, as shown by the Wannier center flow in \ref{fig: Rashba topological phase diagram}(f) of the main text.
The irreps of CB1 and CB2 together do not change and are still $\mathbb Z_2$ non-trivial. 

\subsection{Material realization}
Finally, we discuss the possible material realization of the moir\'e Rashba system.
The Rashba materials are very common in 2D semiconductor heterostructures with electron gases. In certain types of systems, the Rashba strength can be large and easily tuned by the displacement field of external gate voltages. 
Meanwhile, high mobility in 2D semiconductor heterostructures can potentially support large moir\'e unit cell length $L_\text M$ to be comparable to $L_\text R$ for the band-folding mechanism to work.
In Tab.~\ref{tab:Rashba materials} we summarized the promising semiconductor heterostructures with Rashba SOC \cite{balocchi2011full, chin1991high, hayakawa1988enhancement, grundler2000large, PhysRevLett.78.1335, knez2012quantum, luo1990effects, das1989evidence, nguyen1992magneto}. 
Taking the InAs quantum well \cite{grundler2000large} as an example, it has $L_R \sim 96$nm and the corresponding energy scale $E_0 \sim a / L_R^2 \sim 1.1 \text{meV}$. Patterning hole arrays in substrates allow for a superlattice potential with the unit cell length $L_\text M \gtrsim40$nm and tunable superlattice potential strength up to $\Delta_1\sim 10$meV \cite{forsythe2018band, barcons2022engineering}. Thus, the moire length scale $L_\text M$ and the Rashba length $L_R$ can be comparable.
Furthermore, the high mobility in InAs quantum wells guarantees a small disorder broadening $E_\tau \sim 0.15 \text{meV}$, which is much smaller than $E_0$ and $\Delta_1$.


\begin{table}[]
    \centering
    \includegraphics[width=\columnwidth]{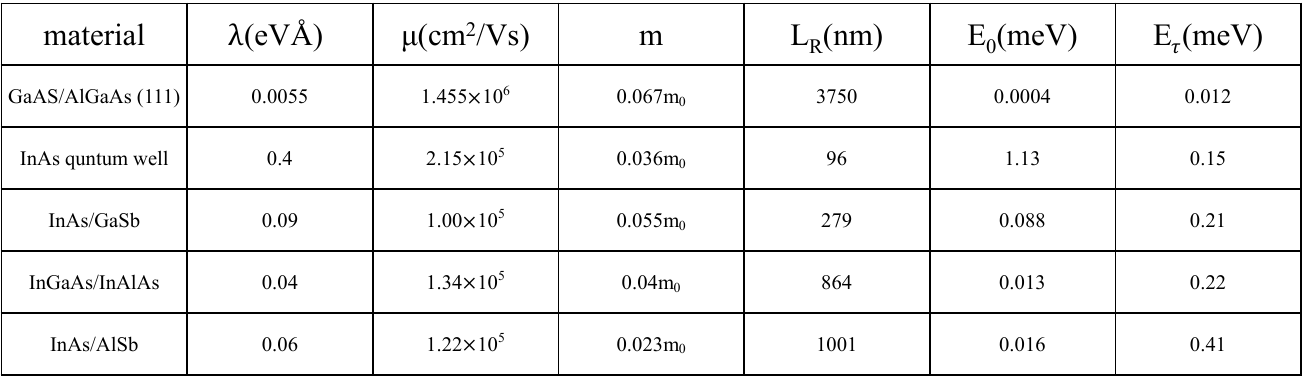}
    \caption{
    Promising Rashba semiconductor heterostructures for the moir\'e Rashba model. 
    $\lambda, \mu, m, L_\text R, E_0, \text{and} \ E_\tau$ are the Rashba SOC, mobility, effective mass, spin precession length, miniband energy scale, and disorder broadening energy scale, respectively. $a=\hbar^2/2m$ in \eqnref{eq: rashba ham} of the main text.
    $E_0 = a (\pi / L_\text R)^2$ is the energy scale for the minibands when the band inversion between spectra folded from 1st and 2nd MBZ happens.
    $E_\tau = e \hbar / \mu m$ is the disorder broadening.
    }
    \label{tab:Rashba materials}
\end{table}

\section{Topological mini-bands in Moir\'e BHZ model}
\label{sec:moire BHZ}
In this section, we will discuss the BHZ model under a moir\'e superlattice potential and study its nontrivial band topology.

\subsection{Moir\'e BHZ model and its symmetry property}
\begin{figure}
\centering
\includegraphics[width=1\textwidth]{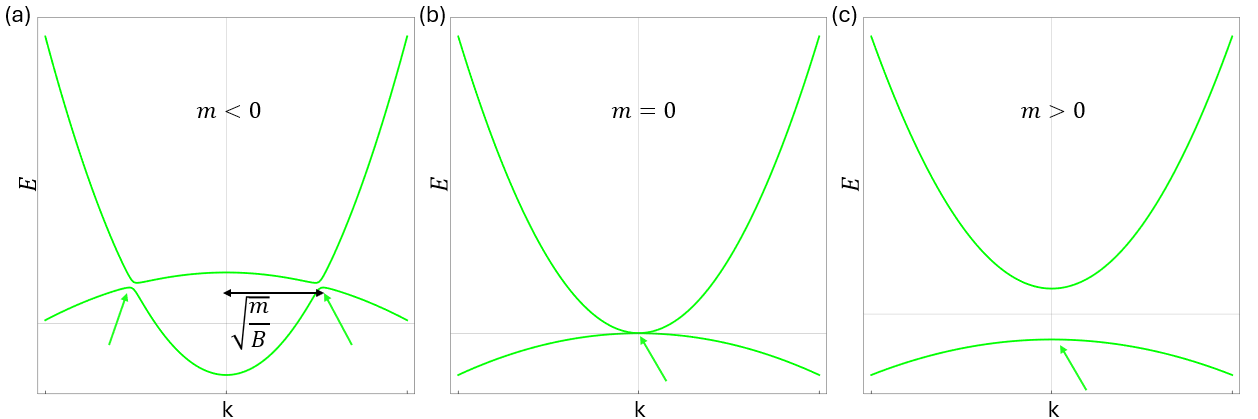}
\caption{\label{fig:band invert} (a)-(c) Band dispersions with $\Delta_1$=0, and $A=0.01 \text{nm}\cdot $eV. Here we choose a very small $A$ value to illustrate why the valence band maximum is away from the $\Gamma$ point. The green arrow points to the maximum of the valence band. When $m>0$, the valence band maximum is at $\Gamma$. When $m=0$, the valence band and conduction band touch at $\Gamma$. When $m<0$, the valence band maximum is at $k=\sqrt{\frac{m}{B}}$ away from $\Gamma$. 
}
\end{figure}

The model  Hamiltonian can be write as following 
\begin{eqnarray}\label{eq: Moire BHZ}
&&H=H_0+H_ \text M+H_R\label{eq:H moire all}\\
&& H_{0}(k) =\epsilon(k)I+\left( \begin{array}{cccc} \mathcal{M}(k)&Ak_+&0&0\\ 
  A k_-& -\mathcal{M}(k)&0&0\\
0 & 0&\mathcal{M}(k)&-Ak_-\\
  0 & 0&-Ak_+&-\mathcal{M}(k)
 \end{array} \right) \label{eq:H BHZ}\\
 && H_{R}(k) =\left( \begin{array}{cccc} 0&0&i \lambda k_-&0\\ 
0& 0&0&0\\
-i \lambda k_+ & 0&0&0\\
  0 & 0&0&0
 \end{array} \right) \label{eq: H rashba}\\
  &&   H _ \text M (\bs r) = \Delta_1 \sum_{\bs g}  e^{i \bs g \cdot \bs r} I
\end{eqnarray}
with $k_\pm=k_x\pm i k_y $, $\mathcal{M}(k)=m-B k^2$, $\epsilon(k)=-Dk^2$, and $I$ is an identity matrix. The basis of above Hamiltonian are $\ket{S,\uparrow}$,$\ket{P_+,\uparrow}$, $\ket{S,\downarrow}$ and  $\ket{-P_-,\downarrow}$, where $P_\pm$ and $S $ represent $P_x\pm P_y$ and $S$ orbitals, respectively. The arrows $\uparrow, \downarrow$ represent spin up and spin down, respectively. The $H_0$ term is the BHZ Hamiltonian \cite{bernevig2006quantum}. $H_\text M(\bs r)$ is the moir\'e potential, the summation of $\bs g$ is as same as that in Eq.(\ref{eq: moire potential main}). The $H_{R}$ term is Rashba SOC term. For the HgTe quantum wells, the material dependent parameters are given by $A=0.365 nm\cdot$ eV, $B=-0.706 nm^2 \cdot$ eV, and $D=-0.532 nm\cdot$ eV\cite{rothe2010fingerprint}.
For the moir\'e potential, we choose $|b^{M}_1|=\frac{\pi^2}{9 \sqrt{3}}nm^{-1}$, so that the moir\'e unit cell length $L_m=11.46nm$. We treat the gap parameter $m$, the Rashba SOC parameter $\lambda$ and the moir\'e potential strength $\Delta_1$ as tuning parameters. When both $\Delta_1=0$ $\lambda$=0 and $m<0$, the valence band maximum can be away from $\Gamma$. In the weak coupling limit $A \rightarrow0$, the valence band maximum is located around $k\sim \sqrt{\frac{m}{B}}$. Fig.\ref{fig:band invert} depicts the evolution of band dispersion for (a) negative, (b) zero and (c) positive $m$ in the $A \rightarrow0$ limit, respectively. 

Below, we first consider Rashba SOC parameter $\lambda=0$ and then discussion the case with nonzero $\lambda$.   

The BHZ Hamiltonian $H_0$ in Eq.(\ref{eq:H BHZ}) has the $D_{\infty h} $ symmetry, including the full rotation about z-axis, the mirror symmetries along the x, y, and z axis, denoted as $\mathcal{M}_x$, $\mathcal{M}_y$ and $\mathcal{M}_z$, inversion symmetry $\mathcal{I}$, as well as the time reversal symmetry $\mathcal{T}$. With the moir\'e potential, the symmetry group of the Hamiltonian $H_0+H_M$ is $P6/mmm$ that can be generated by $\mathcal{M}_y$, $\mathcal{M}_z$, six-fold rotation around the z-axis $\mathcal{C}_{6z}$, as well as $\mathcal{T}$. The inversion $\mathcal{I}$ is also in this space group as $\mathcal{I} = \mathcal{M}_z\mathcal{C}_{6z}^3$. Because of $\mathcal{T}$ and $\mathcal{I}$, all irreps in the group are doubly degenerate. Furthermore, due to the mirror $\mathcal{M}_z$, The Hamiltonian $H_0+H_M$ can be written in a block diagonal with two blocks characterized by the opposite $\mathcal{M}_z$ eigenvalue, namely, $M_z=i$ for the upper block and $M_z=-i$ for the lower block. The upper block Hamiltonian is written as
\begin{eqnarray}
H_{upper}(k) =\left( \begin{array}{cc} \mathcal{M}(k)+\epsilon(k)+\Delta_1 \sum_{\bs g}  e^{i \bs g \cdot \bs r}&Ak_+\\ 
  A k_-& -\mathcal{M}(k)+\epsilon(k)+\Delta_1 \sum_{\bs g}  e^{i \bs g \cdot \bs r}
 \end{array} \right), \label{eq:H upper}
\end{eqnarray}
while the lower block $H_{lower}$ can be related to $H_{upper}$ by $\mathcal{T}$. The symmetry group for $H_{upper}$ is described by magnetic space group P6/mm'm' with the generators $C_{6z}$, $\mathcal I$ and  $\mathcal M_y \mathcal T$. The symmetry group of $H_{upper}$ turns out to be useful for our understanding of topological phase transitions discussed in Sec.B below. 
For the full Hamiltonian $H=H_0+H_M+H_R$, in which the Rashba SOC $H_R$ breaks both mirror $\mathcal{M}_z$ and inversion $\mathcal I$, the symmetry group is further lowered to $P6mm$ with generators $\mathcal{M}_y$, $\mathcal{C}_{6z}$ and $\mathcal{T}$. 

\subsection{Topological phase diagram of the moir\'e BHZ model with inversion symmetry}
In the main text, we have discussed a variety of topological phases with different mirror Chern numbers and the corresponding topological phase transitions between them for the minibands VB1, which are mainly summarized in Fig. 3c of the main text. In particular, we focused on the phase transitions between moir\'e minibands from different momentum shells. In this section, we will provide more details on different topological phases and apply our moir\'e TQC formalism developed in Sec.\ref{sec:moire TQC} to understand the topological phase transition lines of VB1 around the weak moir\'e potential $\Delta_1\rightarrow 0$ limit in Fig. 3c of the main text. 

As discussed in the main text, the topological phase transitions in Fig. 3c involves the band inversion around $\Gamma$ (green and black lines), $M$ (yellow and red lines) and $K$ (white line). These phase transition lines separate different topological phases for VB1, which can be characterized by the irreps of VB1 at high symmetry momenta, denoted by ($\bar{\Gamma}_i,\bar{K}_j,\bar{\text M}_k$) or ($i,j,k$) for short. Below we will discuss these phase transition lines and the corresponding topological phases separately with the help of moir\'e TQC. We focus on the Hamiltonian $H_0+H_M$ without the Rashba SOC $H_R$ so that the inversion symmetry is preserved in this section and will discuss the influence of Rashba SOC $H_R$ that breaks inversion symmetry in the next section.  

\subsubsection{Mapping the effective model in Region A to the Kane-Mele model}
\begin{figure}
\centering
\includegraphics[width=1\textwidth]{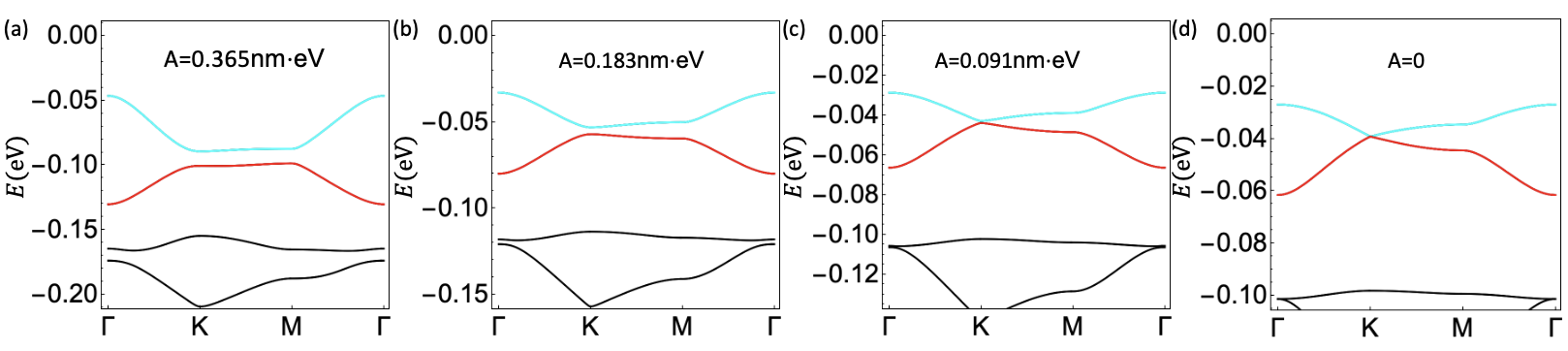}
\caption{\label{fig:reduce A SM} (a)-(d) Band dispersions with m=0.1eV and $\Delta_1=-0.05\text{eV}$. The linear term A decrease from 0.365nm$\cdot$eV to 0. The VB1 and VB2 are highlighted by cyan and right. VB1 is always separately in energy from the
VB2 and the conduction bands when decreasing A from a large positive value to a very small positive one. The mini-gap between the VB1 and VB2 is closed at $K$ when $A=0$. 
}
\end{figure} 
We start with the discussion of the non-trivial mirror Chern number in the region A. As the gap parameter $m$ is positive in this region, the BHZ model $H_0$ is in the trivial phase by itself, and the nontrivial mirror Chern number comes from the influence of the moir\'e superlattice potential $H_M$. As we consider $\Delta_1<0$ in $H_M$, the valence bands feel a honeycomb lattice potential, so we expect our model here can be adiabatically connected to the Kane-Mele model, similar to the case of topological insulator thin film model in a honeycomb mori\'e superlattice studied in Ref.\cite{yang2024topological}. However, we notice that there is a difference in the basis wavefunctions between the BHZ model studied here and the TI thin film model studied in Ref. \cite{yang2024topological}. In the region A, the VB1 and the 2nd valence minibands, denoted as VB2, together form an EBR that corresponds to the orbitals with the angular momentum $J=\pm\frac{3}{2}$ in the honeycomb lattice. In contrast, the atomic orbitals with the angular momentum $J=\pm\frac{1}{2}$ appear in the honeycomb lattice in Ref.\cite{yang2024topological}. 

Non-trivial topology of this model in the regions with $m>0$ and $\Delta_1\rightarrow 0^-$ is guaranteed by the rotation eigen-values at high symmetry momenta $\Gamma$, $K$ and $M$ in MBZ. The irreps of VB1 and VB2 at high symmetry point is ($\bar\Gamma_{10}$,$\bar K_8$,$\bar M_5$) and ($\bar\Gamma_{7}$,$\bar K_9$,$\bar M_6$), respectively. For the $M_z=+i$ block, the irreps of VB1 are ($\bar\Gamma^M_{13}$,$\bar K^M_{12}$,$\bar M^M_4$), where we use $\bar k^M_i$ to label the irrep of P6/m'm'm at high symmetry point $k$. Because of $C_6$ symmetry, the Chern number of the $M_z=+i$ block is guaranteed to be $6n+1$ with an integer $n$ by the relation between the Chern number and the rotation eigen-values\cite{fang2012bulk},
\begin{eqnarray}\label{eq: cm c6}
    e^{i C \pi/3}=- \eta(\Gamma) \theta(K)\epsilon(M),
\end{eqnarray}  
where $\theta(K)$, $\eta(\Gamma)$ and $ \epsilon(M)$ represent the eigenvalues of $\mathcal{C}_{3z}$ at $K$,
$\mathcal{C}_{6z}$ at $\Gamma$ and  $\mathcal{C}_{2z}$ at $M$, respectively.
These rotation eigenvalues are completely determined by the irreps ($\bar\Gamma^M_{13}$,$\bar K^M_{12}$,$\bar M^M_4$) of the $M_z=+i$ block for the VB1, which gives $\eta(\Gamma)=-i$, $\theta(K)=e^{i\frac{\pi}{3}}$ and $\epsilon(M)=-i$ . The values of $\eta, \theta, \epsilon$ for other irreps can be found in the Tab.\ref{fig:charater table gamm}, \ref{fig:charater table M} and \ref{fig:charater table K}. 


To show explicitly the adiabatic connection between our model $H_0+H_M$ and the Kane-Mele model, we decrease the parameter value $A$ for the linear term. As shown in the Fig.\ref{fig:reduce A SM}a-c, the VB1 is always separately in energy from the VB2 and the conduction bands when decreasing $A$ from a large positive value to a very small positive one. In the $A\rightarrow0^+$ limit, we can treat the linear term as a perturbation. At $A=0$ (Fig.\ref{fig:reduce A SM}d), VB1 and VB2 are degenerate at $K$, giving rising to a Dirac cone, similar to the case of the graphene model. With a small $A$, we find the linear term can open a gap of the Dirac cone, leading to the effective model for both the VB1 at VB2 near $K$ as
\begin{eqnarray}\label{eq: H eff at K}
   H^{eff}_K(k)=E_0(K)+\Delta_K \tau_0 \sigma_z+V_f ( k_x \tau_z \sigma_x-k_y \tau_0 \sigma_y)
\end{eqnarray}
on the basis $\ket{-\frac{1}{2},+i}$,$\ket{\frac{1}{2},+i}$,$\ket{\frac{1}{2},-i}$ and $\ket{-\frac{1}{2},-i}$, where the basis function $\ket{J,M_z}$ is labelled by the eigen-values of symmetry operators $\mathcal{C}_{3z}$ and $\mathcal{M}_z$ as
\begin{eqnarray}\label{eq: wavefunction baisi} 
\mathcal{C}_{3z} \ket{J,M_z}= e^{- i\frac{2J \pi}{3}} \ket{J,M_z};\mathcal{M}_{z} \ket{J,M_z}= M_z \ket{J,M_z}.
\end{eqnarray}
$\ket{-\frac{1}{2},+i}$ and $\ket{\frac{1}{2},-i}$ form the irrep $\bar K_8$, while $\ket{-\frac{1}{2},-i}$ and $\ket{\frac{1}{2},i}$ form the irrep $\bar K_9$.The Pauli matrices $\sigma$ and $\tau$ are defined on the basis of $M_z$ and $C_{3z}$ eigen-states, respectively. The parameters in the effective model can be calculated from perturbation theory as 
\begin{eqnarray}
    &&E_0(K)=-D K^2-\sqrt{A^2K^2+\mathcal{M}(K)^2};\nonumber\\
    &&V_f=(-2D+\frac{-A^2+2BM}{\sqrt{A^2K^2+\mathcal{M}(K)^2}})|K|;\nonumber\\
    &&\Delta_K=\frac{(A K)^2}{2}(\frac{1}{-2\sqrt{A^2K^2+\mathcal{M}(K)^2}}-\frac{1}{-2\sqrt{A^2K^2+\mathcal{M}(K)^2}-3\Delta_1})
\end{eqnarray}

The effective Hamiltonian in Eq.(\ref{eq: H eff at K}) exactly takes the form of the Kane-Mele model near $K$.

\subsubsection{Phase transition lines at $\Gamma$ (green and black lines)}


In the phase diagram Fig. 3c of the main text, there are two phase transition lines occurring at $\Gamma$ (green and black lines). These two transitions, although both at $\Gamma$, have different physical origins. The transition of the green line occurs when the gap parameter $m$ reverses its sign in the $\Delta_1\rightarrow0$ limit. Thus, this transition just corresponds to the band inversion of the original BHZ model $H_0$. In the $\Delta_1\rightarrow0$, the irrep of VB1 at $\Gamma$ is completely determined by $H_0$. For $m>0$, the irrep of VB1 is $\bar\Gamma_{10}$, while for $m<0$, the irrep of VB1 is $\bar\Gamma_{9}$. Thus across the green line in the Fig.\ref{fig: bhz phase main}c of the main text, the irrep of VB1 change from $\bar\Gamma_{10}$ in Region A to $\bar\Gamma_{9}$ in Region B, corresponding to a change of the mirror Chern number by 1, from $C_{\mathcal M}=1$ in Region A to $C_{\mathcal M}=2$ in Region B. The typical energy dispersion of mini-bands in Region A and B is shown in Fig.\ref{fig:band SM}a and b, respectively. 

The physical origin of the phase transition across the black line is because of the band inversion between the minibands from the 1st MBZ and those from the 2nd MBZ due to the band folding mechanism, as discussed in the main text. In the following, we will provide a band representation analysis of this phase transition line based on the moir\'e TQC, particularly illustrating the underlying reasons for the change of mirror Chern number by 2 across this transition line. Here we focus on the upper block $H_{upper}$ in the $M_z=+i$ subspace, for which the eigen-states can be described by the irreps denoted by $\tilde \Lambda^M_{k}$ at the certain momentum $k$ for the wave vector groups of magnetic space group $P6/mm'm'$. At $\Gamma$, the wave vector group of $P6/mm'm'$ is labelled by $6/mm'm'$ with its character table shown in the Tab. \ref{fig:charater table gamm} b and c according to the Bilbao Crystallographic Server\cite{aroyo2011crystallography,aroyo2006bilbao,aroyo2006bilbao1}. 

As $6/mm'm'$ is a subgroup of $D_{6h}$, all the irreps of $D_{6h}$ can be decomposed into those of $6/mm'm'$ from the Tab. \ref{fig:charater table gamm} as
\begin{eqnarray}\label{6/mm'm'}
&&\bar{\Gamma}_7=\bar{\Gamma}^M_7\oplus \bar{\Gamma}^M_8;\bar{\Gamma}_8=\bar{\Gamma}^M_9\oplus \bar{\Gamma}^M_{12};\bar{\Gamma}_9=\bar{\Gamma}^M_{10}\oplus \bar{\Gamma}^M_{11};\nonumber\\
&&\bar{\Gamma}_{10}=\bar{\Gamma}^M_{13}\oplus \bar{\Gamma}^M_{14};\bar{\Gamma}_{11}=\bar{\Gamma}^M_{15}\oplus \bar{\Gamma}^M_{18};\bar{\Gamma}_{12}=\bar{\Gamma}^M_{16}\oplus \bar{\Gamma}^M_{17}.
\end{eqnarray}

\begin{table}
\centering
\includegraphics[width=1\textwidth]{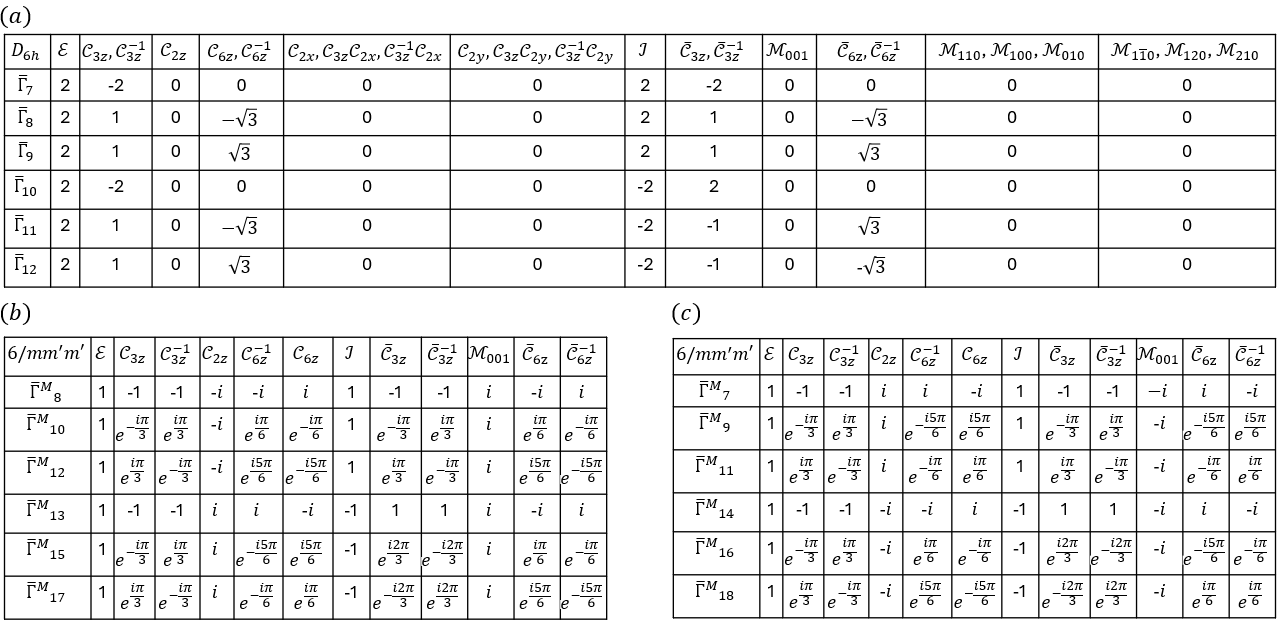}
\caption{\label{fig:charater table gamm} (a) The character table for the wave vector group $D_{6h}$. (b)-(c) The character table of the unitary symmetry operations in wave vector group $6/mm'm'$. In the irrep listed in (b), $M_z=i$, while in the irrep listed in (c),  $M_z=-i$.
}
\end{table}

\begin{figure}
\centering
\includegraphics[width=1\textwidth]{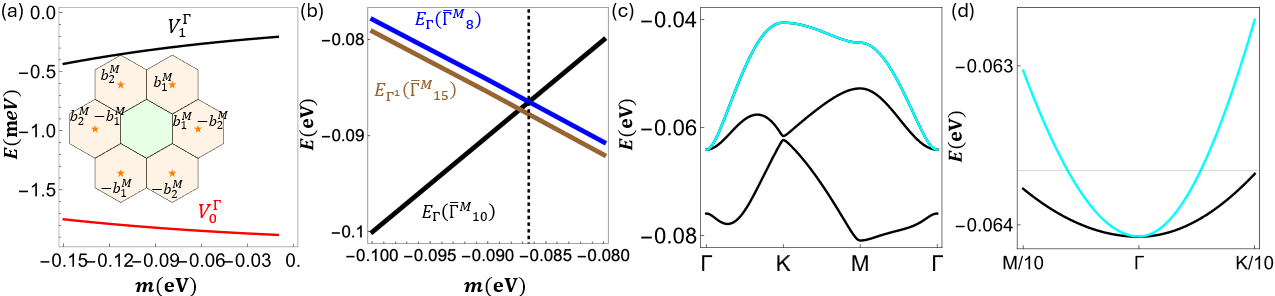}
\caption{\label{fig:TQC gamma} (a) The value of $V^\Gamma_i$ as a function of m at $\Delta_1=-0.002$eV. The insert shows the momenta in the $\tilde A_{\Gamma^1}$ (b) $E_\Gamma(\bar \Gamma^M_{10})$, $E_{\Gamma^1}(\bar \Gamma^M_{15})$ and $E_{\Gamma^1}(\bar \Gamma^M_8)$ near the transition line at $\Delta_1=-0.002$eV (c) Band dispersion at the black line. (d) Band dispersion near the $\Gamma$ point of (c). The band touching at the $\Gamma$ point is quadratic. 
}
\end{figure}

Two minibands with irrep $\bar{\Gamma}^M_i$ and $\bar{\Gamma}^M_j$, which are decomposed from the same $\bar{\Gamma}_i$, have different $M_z$ eigenvalue and are degenerate. For the block $H_{upper}$, only $\bar{\Gamma}^M_8$, $\bar{\Gamma}^M_{10}$, $\bar{\Gamma}^M_{12}$,$\bar{\Gamma}^M_{13}$,$\bar{\Gamma}^M_{15}$ and $\bar{\Gamma}^M_{17}$ are relevant. 


At the momentum shell $\tilde A_\Gamma$ in the 1st MBZ, the four basis wavefunctions of $H_0$ belong to the irreps $\bar{\Gamma}_{10}$ and $\bar{\Gamma}_{9}$, or alternatively $ \bar \Gamma^M_{13}$ and $ \bar \Gamma^M_{10}$ for the upper block $H_{upper}$, respectively. The corresponding eigen-energies in small $\Delta$ limit are 
\begin{eqnarray} \label{eq: E Gamma}
    E_\Gamma( \bar \Gamma^M_{10})=m; \quad E_\Gamma( \bar \Gamma^M_{13})=-m. 
\end{eqnarray}

We next consider the 6 momenta in the momentum shell $\tilde A_{\Gamma^1}$ in the 2nd MBZ and focus on the upper block $H_{upper}$ in Eq.(\ref{eq:H upper}). For zero $\Delta_1$, at each momentum in the ABZ, there are two eigen-states with the eigen-energies 
\begin{eqnarray} \label{eq: eigen energy}
E_\pm(k)=-Dk^2\pm\sqrt{Ak^2+\mathcal{M}^2(k)},
\end{eqnarray}
where $\pm$ represent the conduction and valence bands, respectively. The corresponding eigen-wave functions are labelled as $\ket{\pm,k,M_z=i}$ or abbreviated as $\ket{\pm,k,i}$. Here we only focus on the lower energy eigen-states for the valence bands at 6 momenta in $\tilde A_{\Gamma^1}$, 
$\ket{  -,b^M_1-b^M_2,i}$,$\ket{-,b^M_1,i}$,$\ket{ -,b^M_2,i}$,$\ket{ -,b^M_2-b^M_1,i}$,$\ket{ -,-b^M_1,i}$,$\ket{ -,-b^M_2,i}$, and denote the corresponding representation as $\tilde \Lambda^M_{\Gamma^1}$. 
On the basis of these 6 wave functions, the representation matrices for $\mathcal C_{6z}$, $\mathcal{I}$ and  $M_y \mathcal{T}$ operators have the forms 
\begin{equation}
    \mathcal C_{6z}(\tilde \Lambda^M_{\Gamma^1}) = 
    \begin{pmatrix}
        0 & 0 & 0 & 0 & 0 & -i \\
       -i& 0 & 0 & 0 & 0 & 0 \\
        0 &-i & 0 & 0 & 0 & 0 \\
        0 & 0 & -i & 0 & 0 & 0 \\
        0 & 0 & 0 & -i & 0 & 0 \\
        0 & 0 & 0 & 0 & -i & 0 
    \end{pmatrix};  \mathcal I (\tilde \Lambda^M_{\Gamma^1}) = 
    \begin{pmatrix}
        0 & 0 & 0 & -1 & 0 & 0 \\
       0& 0 & 0 & 0 & -1 & 0 \\
        0 &0 & 0 & 0 & 0 & -1 \\
        -1 & 0 & 0 & 0 & 0 & 0 \\
        0 & -1 & 0 & 0 & 0 & 0 \\
        0 & 0 & -1 & 0 & 0 & 0 
    \end{pmatrix};
    M_y \mathcal T (\tilde \Lambda^M_{\Gamma^1}) = 
    \begin{pmatrix}
        0 & 0 & 0 & 1 & 0 & 0 \\
      0& 0 & 1 & 0 & 0 & 0 \\
        0 &1 & 0 & 0 & 0 & 0 \\
        1 & 0 & 0 & 0 & 0 & 0 \\
        0 & 0 & 0 & 0 & 0 &1 \\
        0 & 0 & 0 & 0 & 1 & 0 
    \end{pmatrix}.
\end{equation}


The representation $\tilde \Lambda^M_{\Gamma^1}$ is reducible, and based on the representation matrices and the character table Tab.\ref{fig:charater table gamm}, we find
\begin{equation}
    \tilde \Lambda^M_{\Gamma^1} = \bar{\Gamma}^M_8\oplus\bar{\Gamma}^M_{10}\oplus\bar{\Gamma}^M_{12}\oplus\bar{\Gamma}^M_{13}\oplus \bar{\Gamma}^M_{15}\oplus \bar{\Gamma}^M_{17}
\end{equation}

By considering the first order perturbation for the Moir\'e potential $H_M$, we derive the form of the effective model on these 6 eigen-state basis as
\begin{equation}
    \mathcal H(\Gamma^1) = 
    \begin{pmatrix}
        E_-(b^M_1)& V^{\Gamma}_0 +  i V^{\Gamma}_1 & 0 & 0 & 0 &  V^{\Gamma}_0 -  i V^{\Gamma}_1 \\
       V^{\Gamma}_0 -  i V^{\Gamma}_1&      E_-(b^M_1) & V^{\Gamma}_0 +  i V^{\Gamma}_1 & 0 & 0 & 0 \\
        0 & V^{\Gamma}_0 -  i V^{\Gamma}_1 &      E_-(b^M_1)& V^{\Gamma}_0 +  i V^{\Gamma}_1 & 0 & 0 \\
        0 & 0 &  V^{\Gamma}_0 - i V^{\Gamma}_1 &       E_-(b^M_1)& V^{\Gamma}_0 +  i V^{\Gamma}_1 & 0 \\
        0 & 0 & 0 &  V^{\Gamma}_0-  i V^{\Gamma}_1 &       E_-(b^M_1) & V^{\Gamma}_0 +  i V^{\Gamma}_1 \\
        V^{\Gamma}_0 +  i V^{\Gamma}_1 & 0 & 0 & 0 &  V^{\Gamma}_0 -  i V^{\Gamma}_1 &     E_-(b^M_1)
    \end{pmatrix},  
\end{equation}
where  $E_-(b^M_1)$ is obtained from Eq.(\ref{eq: eigen energy}). The coupling term $V^\Gamma_i$ is defined as following:
\begin{eqnarray}
    V^{\Gamma}_0=Re\left(\Delta_1 \braket{-,b^M_1,i}{-,b^M_2,i}\right);V^{\Gamma}_1=Im\left(\Delta_1 \braket{-,b^M_1,i}{-,b^M_2,i}\right).
\end{eqnarray}
This Hamiltonian satisfies the symmetry properties, 
\begin{equation}
    C_{6z}(\tilde \Lambda^M_{\Gamma^1})\mathcal H(\Gamma^1) C^{-1}_{6z}(\tilde \Lambda^M_{\Gamma^1})=\mathcal H(\Gamma^1); \mathcal I (\tilde \Lambda^M_{\Gamma^1})\mathcal H(\Gamma^1) \mathcal I^{-1}(\tilde \Lambda^M_{\Gamma^1})=\mathcal H(\tilde \Lambda^M_{\Gamma^1});  M_y \mathcal T (\tilde \Lambda^M_{\Gamma^1}) \mathcal H(\Gamma^1)  (M_y \mathcal T)^{-1}(\tilde \Lambda^M_{\Gamma^1})=\mathcal H^*(\Gamma^1). 
\end{equation}
 Unlike the coupling term in Eq.(\ref{eq:gamma rashba}) that only depend on $\Delta_1$,   $V^\Gamma_0$ and $V^\Gamma_1$  also depends on the parameters $M,B,A,|b^M_1|$.   

The energies of $\mathcal H_{\Gamma^1}$ are
\begin{eqnarray}\label{eq:analytical_eigenenergy_Gamma1}
 E_{\Gamma^1}(\bar \Gamma^M_{17}) =  E(\Gamma^1)-V^{\Gamma}_0+\sqrt{3} V^{\Gamma}_1
 ; E_{\Gamma^1}(\bar \Gamma^M_{15}) =  E(\Gamma^1)-V^{\Gamma}_0-\sqrt{3} V^{\Gamma}_1; E_{\Gamma^1}(\bar \Gamma^M_{8}) =  E(\Gamma^1)-2V^{\Gamma}_0;\nonumber\\ E_{\Gamma^1}\bar (\Gamma^M_{10}) =  E(\Gamma^1)+V^{\Gamma}_0+\sqrt{3} V^{\Gamma}_1
 ; E_{\Gamma^1}(\bar \Gamma^M_{12}) =  E(\Gamma^1)+V^{\Gamma}_0-\sqrt{3} V^{\Gamma}_1; E_{\Gamma^1}(\bar \Gamma^M_{13}) =  E(\Gamma^1)+2V^{\Gamma}_0;
\end{eqnarray}

Fig. \ref{fig:TQC gamma}a shows $V^{\Gamma}_0$ and $V^{\Gamma}_1$ as a function of $m$ at $\Delta_1=-0.002$eV, and Fig. \ref{fig:TQC gamma}b shows the eigen-energies of the $\bar\Gamma^M_{8}$ and $\bar\Gamma^M_{15}$ states in Eq.(\ref{eq:analytical_eigenenergy_Gamma1}) from $\tilde A_{\Gamma^1}$ and the $\bar \Gamma^M_{10}$ states in Eq.(\ref{eq: E Gamma}) from $\tilde A_{\Gamma}$ as a function of the gap parameter $m$. One can see a band crossing between the $\bar \Gamma^M_{10}$ and $\bar \Gamma^M_{8}$ states at $m=-0.0865$eV, so the VB1 belong to the irrep $\bar \Gamma^M_{8}$ for $m<-0.0865$eV and $\bar \Gamma^M_{10}$ for $m>-0.0865$eV. We notice that the $\mathcal C_{6z}$ rotation eigen-values for the irreps $\bar \Gamma^M_{10}$ and $\bar \Gamma^M_{8}$ are $e^{-\frac{i \pi}{6}}$ and $+i$ respectively, so the band touching at the transition point has a quadratic form, as shown in  in the Fig.\ref{fig:TQC gamma}c and d, and the mirror Chern number $C_M$ is changed by 2 according to Eq.(\ref{eq: cm c6}). 

In the above, we have discussed the irreps of VB1 for the upper block Hamiltonian $H_{upper}$, while in the main text, we use the irreps of the full Hamiltonian $H_0+H_M$. The correspondence can be read out from the decomposition forms in Eq.(\ref{6/mm'm'}), from which the decomposition of the irreps for the momentum shell $\tilde A_{\Gamma^1}$ can be given by
\begin{equation}
    \tilde \Lambda_{\Gamma^1} = \bar{\Gamma}_7\oplus\bar{\Gamma}_{8}\oplus\bar{\Gamma}_{9}\oplus\bar{\Gamma}_{10}\oplus \bar{\Gamma}_{11}\oplus \bar{\Gamma}_{12}
\end{equation}
for the full Hamiltonian. Moreover, the transition between  $\bar\Gamma^M_{10}$ and  $\bar\Gamma^M_{13}$ corresponds to that between $\bar\Gamma_{9}$ and $\bar\Gamma_{10}$ for the full Hamiltonian, according to Eq.(\ref{6/mm'm'}). This transition is illustrated by the green line in Fig.\ref{fig: bhz phase main}c. The transition between  $\bar\Gamma^M_{10}$ and  $\bar\Gamma^M_{13}$ corresponds to that between $\bar\Gamma_{9}$ and $\bar\Gamma_{7}$, which is illustrated by the black line in Fig.\ref{fig: bhz phase main}c.

\subsubsection{Phase transition lines at $M$ (yellow and red lines)}
In this section, we analyze the yellow and red transition lines at high-symmetry momentum $M$, which is described by the wave vector group $D_{2h}$. Here we again only consider the upper block Hamiltonian $H_{upper}$ ($M_z=+i$) that is described by the magnetic little group $mm'm'$ with generators  $\mathcal I$, $M_y \mathcal T$ and $M_z$. For the wave vector group $D_{2h}$, the double group irreps are $\bar{M}_5$ and $\bar{M}_6$, which can be decomposed into the irreps of $mm'm'$ as 
\begin{eqnarray}\label{mm'm'}
&&\bar{M}_5=\bar{M}^M_3\oplus \bar{M}^M_4;\bar{M}_6=\bar{M}^M_5\oplus \bar{M}^M_{6}. 
\end{eqnarray}
For $M_z=+i$ sector ($H_{upper}$ block), the relevant irreps are $\bar{M}^M_4$ and $\bar{M}^M_5$. 

\begin{figure}
\centering
\includegraphics[width=1\textwidth]{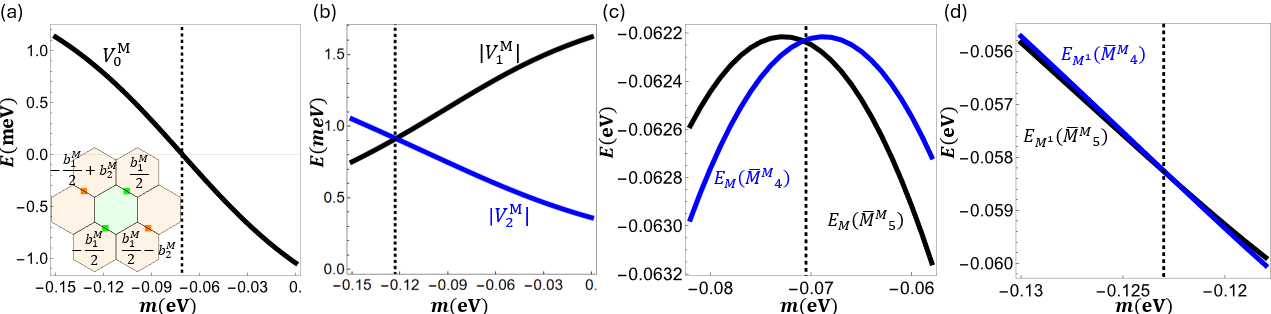}
\caption{\label{fig:TQC M}(a) The value of $V^M_0$ as a function of m at $\Delta_1=-0.002$eV. The insert shows the momenta in $\tilde A_M$ and $\tilde A_{M^1}$ (b)  The value of $|V^M_1|$ and $|V^M_2|$ as a function of m at $\Delta_1=-0.002$eV. (c)  $E_M(\bar M^M_{4})$, $E_M(\bar M^M_{4})$  near the yellow line at $\Delta_1=-0.002$eV (d)  $E_M(\bar M^M_{4})$, $E_M(\bar M^M_{4})$  near the red line at $\Delta_1=-0.006$eV
}
\end{figure}

We label the 1st and 2nd momentum shells for $M$ as $\tilde A_M$ and $\tilde A_{M^1}$, respectively, which are given by  
\begin{eqnarray}  \label{eq:momentum_shell_Gamma}
&&\tilde A_{M} = \{b^M_1/2, -b^M_1/2 \} \nonumber\\
&&\tilde A_{M^1} = \{b^M_1/2-b^M_2, -b^M_1/2+b^M_2 \},
\end{eqnarray}
as shown in the inset of Fig.\ref{fig:TQC M}a.

Here we consider the effective model on both momentum shells $\tilde A_M$ and $\tilde A_{M^1}$ together, and the corresponding basis wave functions are labelled by $\ket{-,b^M_1/2,i}$ and $\ket{ -,-b^M_1/2,i}$ for the irrep $\tilde \Lambda_M$ at $\tilde A_M$ and $\ket{ -,-b^M_1/2+b^M_2,i}$ and $\ket{-, b^M_1/2-b^M_2,i}$ for the irrep $\tilde \Lambda_{M^1}$ at $\tilde A_{M^1}$. The matrix representations of the $\mathcal{I}$ and  $M_y \mathcal{T}$ operators at these basis wave functions can be written as
\begin{equation}
     \mathcal I (\tilde \Lambda^M_{M}\oplus\tilde \Lambda^M_{M^1}) = 
    \begin{pmatrix}
         0 & -1 & 0 & 0 \\
       -1 & 0 & 0 & 0 \\
         0 & 0 & 0 & -1 \\
         0 & 0 & -1 & 0 
    \end{pmatrix};
    M_y \mathcal T (\tilde \Lambda^M_{M}\oplus\tilde \Lambda^M_{M^1}) = 
   \begin{pmatrix}
         0 & 1 & 0 & 0 \\
       1 & 0 & 0 & 0 \\
         0 & 0 & 1 & 0 \\
         0 & 0 & 0 & 1 
    \end{pmatrix}.
\end{equation}
The representation $\tilde \Lambda^M_{M^1}$  and $\tilde \Lambda^M_{M}$ are reducible, and can be decomposed into the irreps of the magnetic little group $mm'm'$ as
\begin{equation}
     \tilde \Lambda^M_{M} = \bar{M}^M_4\oplus\bar{M}^M_5; 
     \tilde \Lambda^M_{M^1} =\bar{M}^M_4\oplus\bar{M}^M_5,
\end{equation}
based on the representation matrices and the character table in Tab.\ref{fig:charater table M}.


By considering the lowest order perturbation for the Moir\'e potential $H_M$, we derive the form of the effective model 
on these 4 eigen-states as
\begin{equation}
    \mathcal H(M\oplus M^1) = 
    \begin{pmatrix}
        E_-(b^M_1/2)& V^M_0 & V^M_1 +  i V^M_2 & V^M_1 -  i V^M_2 \\
       V^M_0&      E_-(b^M_1/2) & V^M_1 -  i V^M_2 & V^M_1 +  i V^M_2  \\
        V^M_1 -  i V^M_2 & V^M_1 +  i V^M_2 &       E_-(b^M_1/2-b^M_2) & 0 \\
        V^M_1 +  i V^M_2 & V^M_1 - i V^M_2 & 0&      E_-(b^M_1/2-b^M_2)
    \end{pmatrix};  
\end{equation}
where 
\begin{eqnarray}
    &&V^{M}_0=\Delta_1 \braket{-,-b^M_1/2,i}{-,b^M_1/2,i}, \nonumber\\
    &&V^{M}_1=Re\left(\Delta_1 \braket{-,b^M_1/2-b^M_2,i}{-,b^M_1/2,i}\right); 
    V^{M}_2=Im\left(\Delta_1 \braket{-,b^M_1/2-b^M_2,i}{-,b^M_1/2,i}\right).
\end{eqnarray}
This Hamiltonian satisfies the symmetry requirements
\begin{eqnarray}
     &&\mathcal I (\tilde \Lambda^M_{M}\oplus\tilde \Lambda^M_{M^1})\mathcal H(M\oplus M^1) \mathcal I^{-1}(\tilde \Lambda^M_{M}\oplus\tilde \Lambda^M_{M^1})=\mathcal H(M\oplus M^1);\\
     &&M_y \mathcal T\mathcal (\tilde \Lambda^M_{M}\oplus\tilde \Lambda^M_{M^1}) H(M\oplus M^1)  (M_y \mathcal T)^{-1}(\tilde \Lambda^M_{M}\oplus\tilde \Lambda^M_{M^1})=\mathcal H^*(M\oplus M^1).
\end{eqnarray}

The eigen-energies of $\mathcal H_{M}$ up to the lowest order perturbation are
\begin{eqnarray} \label{eq: TQC energy M}
 E_{M}(\bar M^M_{5}) =  E_-(b_1^M/2)+V_0^M
 ; E_{M^1}(\bar M^M_{5}) =  E_-(b_1^M/2-b_2^M)+\frac{(V^M_1)^2}{E_-(b_1^M/2)-E_-(b_1^M/2)}\nonumber\\ 
 E_{M}(\bar M^M_{4}) =  E_-(b_1^M/2)-V_0^M
 ; E_{M^1}(\bar M^M_{4}) =  E_-(b_1^M/2-b_2^M)+\frac{(V^M_2)^2}{E_-(b_1^M/2)-E_-(b_1^M/2)},
\end{eqnarray}
where the energy $E_-(k)$ is obtained from Eq.(\ref{eq: eigen energy}) for the momentum $k$. 

For a positive or small negative $m$ ($m>-0.104$eV), we find $E(M)>E(M^1)$ and the VB1 comes from the first momentum shell $\tilde A_{M}$ and thus we only need to compare the eigen-energies $E_{M}(\bar M^M_{5})$ and $E_{M}(\bar M^M_{4})$ in Eq.(\ref{eq: TQC energy M}). The sign of $V^M_0$ depends on $m$ and reverses from negative to positive at $m=-0.071$eV when decreasing $m$, as shown in the Fig.\ref{fig:TQC M}a. Thus, the VB1 carries the irrep $\bar M^M_{4}$ for $m>-0.071$eV and the irrep $\bar M^M_{5}$ for $m<-0.071$eV, following the Eq.(\ref{eq: TQC energy M}) and shown in Fig. \ref{fig:TQC M}c. 


When $m<-0.104$eV, the minibands from the second momentum shell $\tilde A_{M^1}$ have lower energy, so the irrep of VB1 at $M$ depends on the miniband eigen-energies of $E_{M^1}(\bar M^M_{5})$ and $E_{M^1}(\bar M^M_{4})$. Fig. \ref{fig:TQC M}b depicts the amplitudes $|V^{M}_2|$ and $|V^{M}_1|$ as a function of $m$ at $\Delta_1=-0.002$eV, in which we find $|V^{M}_2<|V^{M}_1|$ for $m>-0.123$eV and $|V^{M}_2>|V^{M}_1|$ for $m<-0.123$eV. Therefore, the irrep for the VB1 is also changed at $m=-0.123$eV according to Eq.(\ref{eq: TQC energy M}). It belongs to $\bar M^M_{5}$ for $m>-0.123$eV and $\bar M^M_{4}$ for $m<-0.123$eV, as shown in Fig. Fig. \ref{fig:TQC M}d. 

Combining all the discussions above, we find the VB1 belongs to the irrep $\bar M^M_{5}$ for -0.123eV$<m<$-0.071eV and $\bar M^M_{4}$ for $m>-0.071$eV or  $m<-0.123$eV. 
 
The above analysis of irreps can be extended to the full Hamiltonian. According to Eq.(\ref{mm'm'}), the irreps of the eigen-states in the momentum shells $\tilde A_M$ and $\tilde A_{M^1}$ for the full Hamiltonian are 
\begin{equation}
    \tilde \Lambda_{M} = \bar{M}_5\oplus\bar{M}_{6}; \tilde \Lambda_{M_1} =\bar{M}_5\oplus\bar{M}_{6}. 
\end{equation} 
According to Eq.(\ref{mm'm'}), the transition between  $\bar M^M_{5}$ and  $\bar M^M_{4}$ corresponds to that between $\bar M_{6}$ and $\bar M_{5}$ for the full Hamiltonian. The transition near $m=-0.071$eV ($m=-0.123$eV) corresponds to the yellow (black) line in Fig.\ref{fig: bhz phase main}c of the main text. It is noted that both two transitions are between $\bar M_{6}$ and $\bar M_{5}$ are within the same momentum shell. 
 
\begin{table}
\centering
\includegraphics[width=1\textwidth]{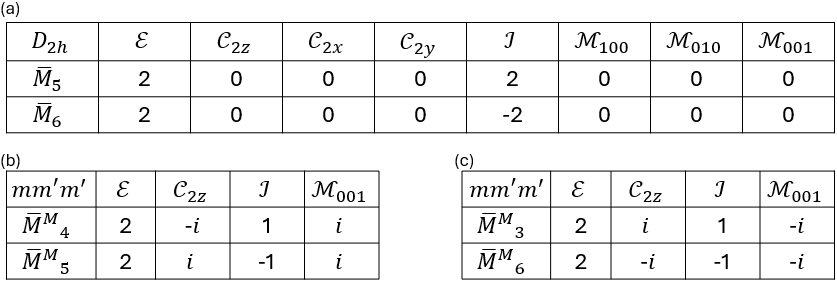}
\caption{\label{fig:charater table M} (a) The character table for the wave vector group $D_{2h}$. (b)-(c) The character table of the unitary symmetry operations in wave vector group $mm'm'$. In the irrep listed in (b), $M_z=i$, while in the irrep listed in (c),  $M_z=-i$.
}
\end{table} 

\subsubsection{Phase transition line at $K$ (white line)}

\begin{table}
\centering
\includegraphics[width=1\textwidth]{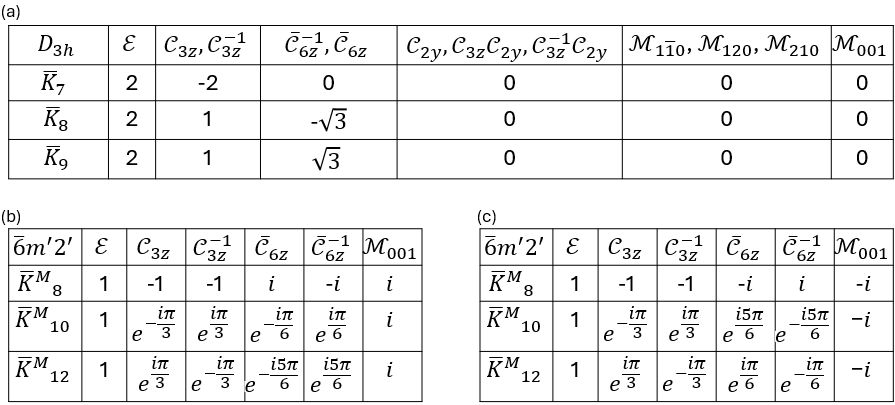}
\caption{\label{fig:charater table K}  (a) The character table for the wave vector group $D_{3h}$. (b)-(c) The character table of the unitary symmetry operations in wave vector group $\bar{6}m'2'$. In the irrep listed in (b), $M_z=i$, while in the irrep listed in (c),  $M_z=-i$.
}
\end{table}

\begin{figure}
\centering
\includegraphics[width=1\textwidth]{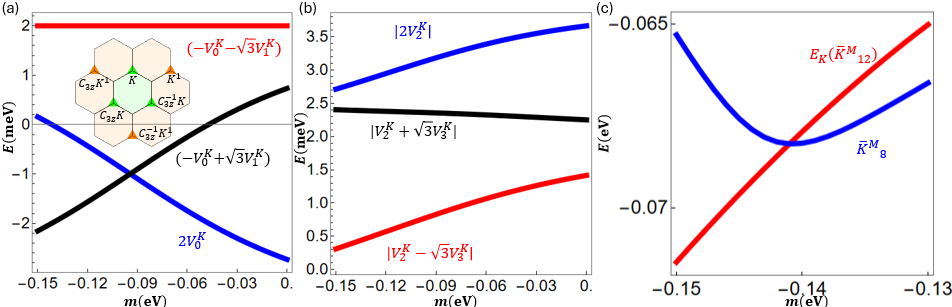}
\caption{\label{fig:TQC K} (a) The value of 2$V^K_0$, $-V^K_0+\sqrt{3}V_1^K$ and $-V^K_0-\sqrt{3}V_1^K$  as a function of m at $\Delta=-0.002$eV. The insert shows the momentum in the $\tilde A_K$ and $\tilde A_{K^1}$ (b) The value of 2$|V^K_2|$, $|V^K_2+\sqrt{3}V_3^K|$ and $|V^K_2-\sqrt{3}V_3^K|$  as a function of m at $\Delta=-0.002$eV. (c) The energy of $\bar K^M_{12}$, and $\bar K^M_{8}$ states near the transition line at $\Delta_1=-0.002$eV.
}
\end{figure} 

In this section, we analyze the white transition lines at high-symmetry momentum $K$ in Fig.\ref{fig: bhz phase main}c of the main text. The wave vector group at $K$ is described by the point group $D_{3h}$ for the full Hamiltonian $H_0+H_M$, and the upper block Hamiltonian $H_{upper}$ ($M_z=+i$) is described by the magnetic little group $\bar{6}m'2'$ with generators $M_z$, $M_y T $ and $\mathcal{I} C_{6z}$. For the wave vector group $D_{3h}$, the double group irreps are $\bar{K}_7$, $\bar{K}_8$ and $\bar{K}_9$, which can be decomposed into the irreps of $\bar{6}m'2'$ as
\begin{eqnarray}\label{-6m'2'}
&&\bar{K}_7=\bar{K}^M_7\oplus \bar{K}^M_8;\bar{K}_8=\bar{K}^M_9\oplus \bar{K}^M_{12};\bar{K}_9=\bar{K}^M_{10}\oplus \bar{K}^M_{11}.
\end{eqnarray}
For $M_z=+i$ sector ($H_{upper}$ block), the relevant irreps are $\bar{K}^M_8$, $\bar{K}^M_{10}$ and $\bar{K}^M_{12}$. 

Here we consider the effective model on both momentum shells $\tilde A_K$ and $\tilde A_{K^1}$ in Eq.(\ref{eq:momentumShell_K}) together, and the corresponding basis wave functions are labelled by $\ket{-,K,i}$, $\ket{ -,C_{3z}K,i}$ and $\ket{ -,C^{-1}_{3z}K,i}$ for the irrep $\tilde \Lambda_K$ at $\tilde A_K$ and $\ket{-,C^{-1}_{3z}K^1,i}$, $\ket{ -,K^1,i}$ and $\ket{ -,C_{3z}K^1,i}$ for the irrep $\tilde \Lambda_{K^1}$ at $\tilde A_{K^1}$.


The matrix representations of $\mathcal {I} C_{6z}$ $M_y \mathcal{T}$ operators on these basis can be written as 
\begin{equation}
    \mathcal {I} C_{6z}(\tilde \Lambda_{K}\oplus\tilde \Lambda_{K^1}) = 
    \begin{pmatrix}
        0 & i & 0 & 0 & 0 & 0 \\
       0& 0 & i & 0 & 0 & 0 \\
        i &0 & 0 & 0 & 0 & 0 \\
        0 & 0 & 0 & 0 & i & 0 \\
        0 & 0 & 0 & 0 & 0 & i \\
        0 & 0 & 0 & i & 0 & 0 
    \end{pmatrix}; 
    M_y \mathcal T (\tilde \Lambda_{K}\oplus\tilde \Lambda_{K^1}) = 
    \begin{pmatrix}
        1 & 0 & 0 & 0 & 0 & 0 \\
      0& 0 & 1 & 0 & 0 & 0 \\
        0 &1 & 0 & 0 & 0 & 0 \\
        0 & 0 & 0 & 1 & 0 & 0 \\
        0 & 0 & 0 & 0 & 0 &1 \\
        0 & 0 & 0 & 1 & 0 & 0 
    \end{pmatrix}.
\end{equation}

From the representation matrices and the character table Tab.\ref{fig:charater table K}, we find the representation $\tilde \Lambda^M_{K}$  and $\tilde \Lambda^M_{K^1}$ can be decomposed by
\begin{equation}
     \tilde \Lambda^M_{K} = \bar{K}^M_8\oplus\bar{K}^M_{10}\oplus\bar{K}^M_{9}; 
    \tilde \Lambda^M_{K^1} = \bar{K}^M_8\oplus\bar{K}^M_{10}\oplus\bar{K}^M_{9}. 
\end{equation}

The effective model 
on these eigen-state basis as is derived as
\begin{equation}
    \mathcal H(K\oplus K^1) = 
    \begin{pmatrix}
        E_-(K)& V^{K}_0 +  i V^{K}_1 & V^{K}_0 -  i V^{K}_1 & 0 & V^{K}_2 -  i V^{K}_3  &  V^{K}_2 +  i V^{K}_3 \\
       V^{K}_0 -  i V^{K}_1 &      E_-(K) & V^{K}_0 +  i V^{K}_1 & V^{K}_2 +  i V^{K}_3  & 0 & V^{K}_2 -  i V^{K}_3  \\
        V^{K}_0 +  i V^{K}_1  & V^{K}_0 -  i V^{K}_1 &      E_-(K) & V^{K}_2 -  i V^{K}_3  & V^{K}_2 +  i V^{K}_3  & 0 \\
        0 & V^{K}_2 -  i V^{K}_3 &  V^{K}_2 +  i V^{K}_3 &      E_-(K^1) & 0& 0 \\
         V^{K}_2 +  i V^{K}_3  & 0 &  V^{K}_2 -  i V^{K}_3  &  0 &      E_-(K^1) & 0 \\
         V^{K}_2 -  i V^{K}_3 &  V^{K}_2 +  i V^{K}_3  & 0 & 0 &  0 &     E_-(K^1)
    \end{pmatrix}.
\end{equation}
where  
\begin{eqnarray}
    &&V^{K}_0=\Delta_1 Re\braket{-,K,i}{-,C^{-1}_{3z}K,i};V^{K}_1= \Delta_1 Im\braket{-,K,i}{-,C^{-1}_{3z}K,i},\nonumber\\ 
     &&V^{K}_2=\Delta_1 Re\braket{-,K^1,i}{-,K,i};V^{K}_3=\Delta_1 Im\braket{-,K^1,i}{-,K,i},
\end{eqnarray}
and the energies $E_-(K)$ and $E_-(K^1)$ can be obtained from Eq.(\ref{eq: eigen energy}). This Hamiltonian satisfies 
\begin{eqnarray}
   &&\mathcal{I} C_{6z}(\tilde \Lambda_{K}\oplus\tilde \Lambda_{K^1})\mathcal H(K\oplus K^1) (\mathcal{I} C_{6z})^{-1}(\tilde \Lambda_{K}\oplus\tilde \Lambda_{K^1})=\mathcal H(K\oplus K^1); \\
   &&M_y \mathcal T (\tilde \Lambda_{K}\oplus\tilde \Lambda_{K^1})\mathcal H(K\oplus K^1)  (M_y \mathcal T)^{-1}(\tilde \Lambda_{K}\oplus\tilde \Lambda_{K^1})=\mathcal H^*(K\oplus K^1).
\end{eqnarray}
The eigen-energies of the above effective Hamiltonian $\mathcal H_{K}$ can be derived perturbatively as
 \begin{eqnarray} \label{eq: tqc energy k}
 &&E_{K}(\bar K^M_{8}) =   E_-(K)+2V_0^K
 ; E_{K^1}(\bar K^M_{8}) =  E_-(K^1)+\frac{(2V^K_2)^2}{E_-(K^1) -E_-(K)}
 , \nonumber\\
&& E_{K}(\bar K^M_{10}) =  E_-(K)-V_0^K+\sqrt{3} V^K_1; E_{K^1}(\bar K^M_{10}) =  E_-(K^1)+\frac{(V_2^K+\sqrt{3} V^K_3)^2}{E_-(K^1) -E_-(K)},\nonumber\\
 &&E_{K}(\bar K^M_{12}) = E_-(K) -V_0^K-\sqrt{3} V^K_1;\bar E_{K^1}(K^M_{12}) =  E_-(K^1)+\frac{(V_2^K-\sqrt{3} V^K_3)^2}{E_-(K^1) -E_-(K)},
\end{eqnarray}
where the energy $E_-(k)$ is obtained from Eq.(\ref{eq: eigen energy}) for the momentum $k$.

When $E_-(K)> E_-(K^1)$, we compare three eigen-energies $E_{K}(\bar K^M_{8})$, $E_{K}(\bar K^M_{10})$ and $E_{K}(\bar K^M_{12})$. Fig.\ref{fig:TQC K}a shows  2$V^K_0$, $-V^K_0+\sqrt{3}V_1^K$ and $-V^K_0-\sqrt{3}V_1^K$ as a function of $m$ at $\Delta=-0.002$eV. From Eq. (\ref{eq: tqc energy k}), we always find the $\bar K^M_{12}$ state has largest energy. 

When band inversion of minibands between the first and 2nd momentum shells happens at $K$ point ($E_-(K)< E_-(K^1)$) for $m<-0.142$eV, the irrep of VB1 at K depend on the energy ordering of $E_{K^1}(\bar K^M_{8})$, $E_{K^1}(\bar K^M_{10})$ and $E_{K^1}(\bar K^M_{12})$. Fig.\ref{fig:TQC K}b shows  2$|V^K_2|$, $|V^K_2+\sqrt{3}V_3^K|$ and $|V^K_2-\sqrt{3}V_3^K|$  as a function of $m$ at $\Delta=-0.002$eV. From Eq. (\ref{eq: tqc energy k}), we find $E_{K_1}(\bar K^M_{8})$ is the largest one. Thus, the band crossing occurs between the $\bar K^M_{8}$ states from $\tilde A_{K^1}$ and the $\bar K^M_{12}$ states from $\tilde A_{K}$ at $m=-0.142$eV, as shown in Fig.\ref{fig:TQC K}c. The VB1 belongs to the irrep $\bar K^M_{8}$ for $m<-0.142$eV and $\bar K^M_{12}$ for $m>-0.142$eV.

The above analysis of irreps can be extended to the full Hamiltonian. 
According to Eq.(\ref{-6m'2'}), the irreps at the momentum shells $\tilde A_{K}$ and $\tilde A_{K^1}$ for the full Hamiltonian are
\begin{equation}
    \tilde \Lambda_{K} = \bar{K}_7\oplus\bar{K}_{8}\oplus\bar{K}_{9}; 
    \tilde \Lambda_{K^1} = \bar{K}_7\oplus\bar{K}_{8}\oplus\bar{K}_{9}. 
\end{equation}
The transition between  $\bar K^M_{8}$ and $\bar K^M_{12}$ corresponds to that between $\bar K_{7}$ and $\bar K_{8}$ for the full Hamiltonian, according to Eq.(\ref{-6m'2'}), which corresponds to the white line in Fig.\ref{fig: bhz phase main}c of the main text.

\subsubsection{Miniband dispersion and topologically nontrivial flat minibands}
\begin{figure}
\centering
\includegraphics[width=1\textwidth]{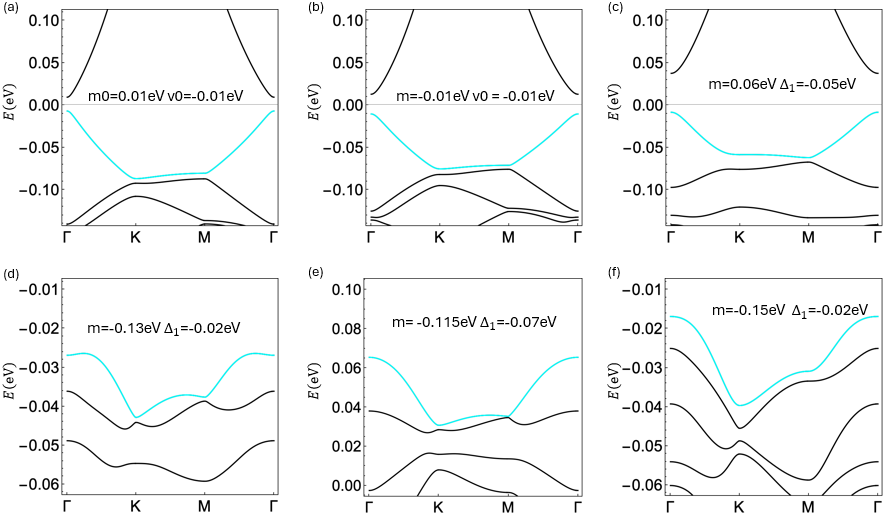}
\caption{\label{fig:band SM} (a)-(c) band dispersions in the region A-C in the Fig.4C in the main text. (d)-(f) band dispersions in the region F-H in the Fig.4C in the main text.
}
\end{figure}

In the main text, we have shown the energy dispersion of the minibands in the region D and E to illustrate the band inversion at $\Gamma$ due to the band-folding mechanism. In Fig.\ref{fig:band SM}, we show the typical energy dispersion for the minibands, particularly the VB1, in the other regions in the phase diagram (Fig. 3c of the main text). Fig.\ref{fig:band SM}a-c correspond region A-C, and Fig.\ref{fig:band SM}d-f correspond region F-H in Fig.3c of the main text.

\begin{figure}
\centering
\includegraphics[width=1\textwidth]{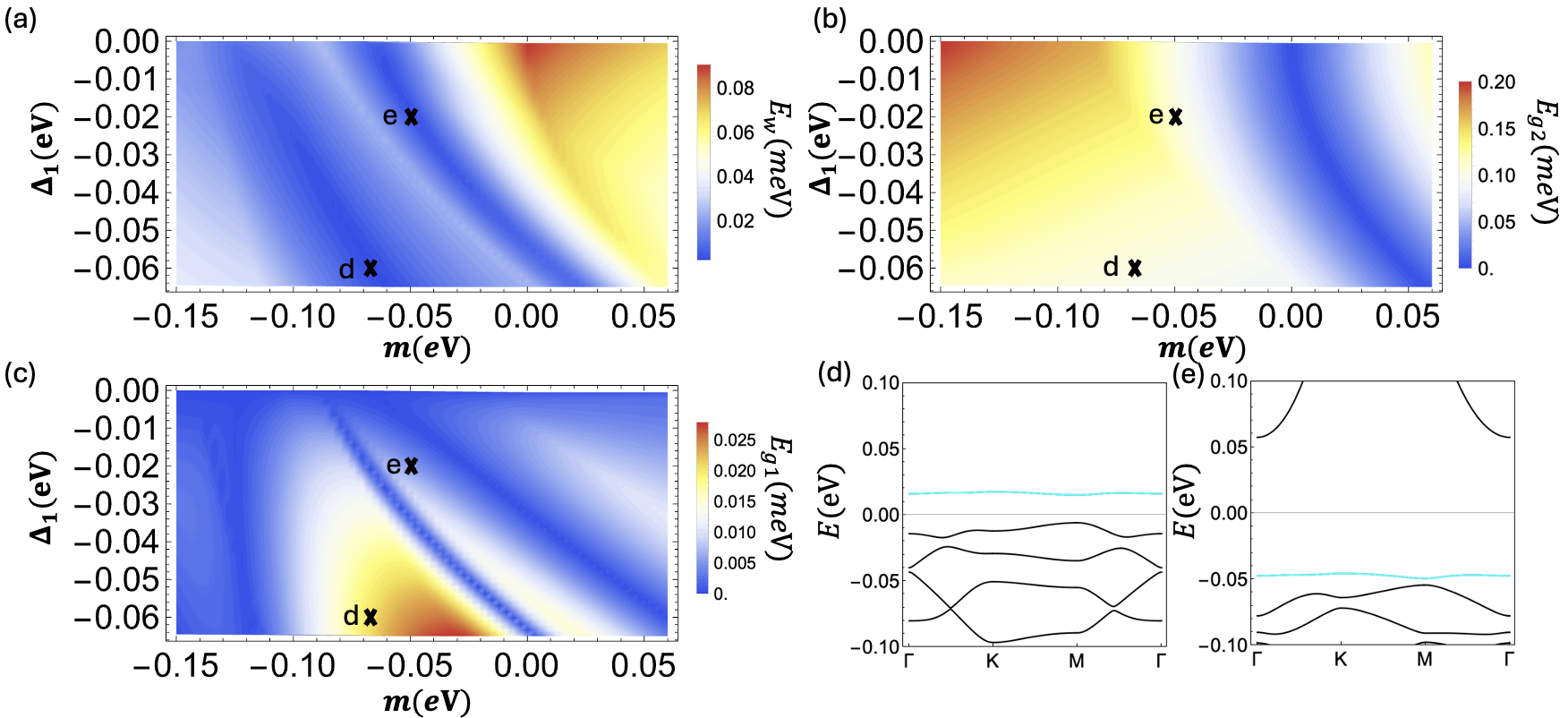}
\caption{\label{fig:gap and width S} (a) Bandwidth $E_w$ of VB1 as function of $\Delta_1$ and $m$ . (b)The band gap between VB1 and CB1 $E_{g2}$  as function of $\Delta_1$ and $m$  . (c) The band gap between VB1 and VB2  $E_{g1}$  as function of $\Delta_1$ and $m$ . (d)-(e) The band dispersion with parameters labeled by the cross in a. 
}
\end{figure}

Generally, we find the minibands of VB1 are dispersive and exhibit a small energy mini-gap from other minibands. To extract more energy dipsersion information for VB1, we depict the bandwidth of the VB1, the mini-band gap between the VB1 and the conduction minibands, and the mini-band gap between the VB1 and VB2 in Fig.\ref{fig:gap and width S}(a), (b) and (c), respectively. An interesting observation is that at the crosses d and e labelled in Fig.\ref{fig:gap and width S}(a), the bandwidth of VB1 is locally minimized while a significant mini-band gap exists between the VB1 and VB2, as well as between the VB1 and conduction minibands. The energy dispersion for the VB1 is depicted in Fig.\ref{fig:gap and width S}(d) and (e) for the parameters at the crosses d and e in \ref{fig:gap and width S}(a). The mirror Chern numbers of the VB1 for the point d and e are -1 and 1, respectively. Thus, the topologically nontrivial flat minibands of the VB1 in these two regions are identified and can potentially support fractional Chern insulator phase. This was very recently discussed in details in  Ref.\cite{tan2024designing}.

\subsection{Topological phase diagram of the moir\'e BHZ model without inversion symmetry}
\subsubsection{Topological phase diagram without inversion}

\begin{figure}
\centering
\includegraphics[width=1\textwidth]{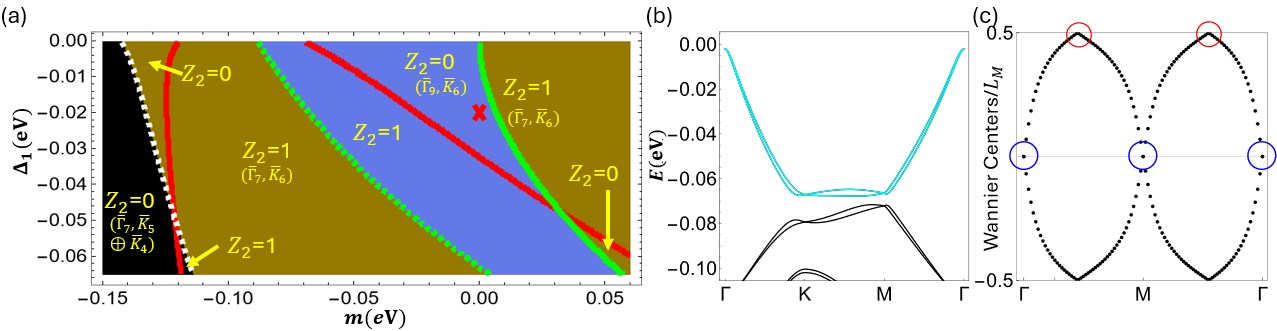}
\caption{\label{fig:phase no I} (a)Topological phase diagram of VB1 without inversion symmetry. 
Different color distinguish irreps at high symmetry points. 
The  solid lines identify $Z_2$ number change. The  green solid lines separate regions with different irreps.   The dashed lines identify irrep change meanwhile $Z_2$ number does not change. (a) Band dispersion with Rashaba SOC strength $\lambda$=0.02eV. and parameters labeled by the red cross in (a). (b) Wannier center flow for VB1. The Wannier center flow still reveals a double winding feature when Rashba SOC is included. The crossings labeled by blue circles at $\Gamma$ and $M$ are protected by  $\mathcal{T}$ and the crossings labeled by red circles are protected by $C_{2z} \mathcal{T}$ symmetry.}
\end{figure} 
In this section, we will consider the topological phase diagram for the VB1 with the full Hamiltonian $H$ in Eq. (\ref{eq: Moire BHZ}) without inversion. We choose a small Rashba SOC term so that there is no additional band closing between VB1 and other bands. As the Rashba SOC term reduces the symmetry group from $P6/mmm$ to $P6mm$, the mirror Chern number is no longer well defined. However, due to the existence of time reversal symmetry, $Z_2$ number is still valid, which can be connected to the mirror Chern number by $\nu=C_{\mathcal M}$ mod 2. Therefore, all the phases with odd mirror Chern number for the VB1 in the phase diagram of Fig. 3c in the main text will also be $Z_2$ non-trivial in the phase diagram after breaking the inversion as shown in Fig.\ref{fig:phase no I}(a). We also label the band irreps of VB1 at high symmetry momenta, which can be obtained from the compatibility relations discussed in the character table in Tab.\ref{tab: irreps at G},\ref{tab: irreps at K},\ref{fig:charater table gamm}-\ref{fig:charater table K}. In this figure, two green lines and one white dashed line, separate regions with different irreps. Those three lines correspond to the green, black and white solid lines in the Fig.3c of the main text. However, the irreps changes are not always accompanied by a $Z_2$ number change. Among these three lines, the $Z_2$ number is only change across the solid green line. In addition, there are another two solid red lines in Fig.\ref{fig:phase no I}(a) that changes the $Z_2$ number although the band irreps do not change across these lines. These two red solid lines correspond to yellow and red solid lines in the Fig.3c of the main text. 

For the $C_{\mathcal M}=\pm 2$ phase in the regions B, C and F of Fig.3c of the main text, the $Z_2$ number is 0, which suggests these two regions should be trivial after taking into account the Rashba SOC term $H_R$. However, as we will discuss below, this is {\it not} true and the model shows the fragile topology that is protected by the $C_{2z}T$ symmetry in these two regions after including the Rashba SOC term that breaks inversion.

\subsubsection{Fragile topology in the regions B}
We take the region B as an example. The band dispersion and the Wannier center flow for the VB1, denoted as $\theta(k)$, are shown in Fig.\ref{fig:phase no I}b and c, respectively. The Wannier center flow $\theta(k)$ for the VB1 reveals a double winding feature, implying the non-trivial band topology in this region. To illustrate the stability of this double winding feature, we examine the symmetry protection of the crossing points in the Wanner center flow, as depicted by two red and three blue circles in Fig.\ref{fig:phase no I}c. The crossing points at three blue circles are at time-reversal invariant momenta $\Gamma$ and $M$, and thus protected by the $\mathcal{T}$ symmetry. The crossings at two red circles are stabilized by the $C_{2z}\mathcal{T}$ symmetry. The $C_{2z}\mathcal{T}$ symmetry operator does not change the momentum $k$, but will reverse the Wannier center $\theta$. Thus, the $C_{2z}\mathcal{T}$ symmetry requires the spectrum of Wannier center flow to be symmetric with respect to $\theta=0$ (or equivalently $\theta=0.5$). Furthermore, $C_{2z}\mathcal{T}$ guarantees that the crossings at $\theta=1/2$, as depicted by the red circles in Fig.\ref{fig:phase no I}c, can only move in the $\theta=1/2$ line, but not be gapped. To see that, we consider the Wilson Hamiltonian, defined by $H_W=\frac{1}{2\pi i} log\left( \mathcal{W}(k)\right)$, where $\mathcal{W}(k)$ is the Wilson loop \cite{alexandradinata2014wilson}. The eigen-values of $H_W(k)$ describes the Wannier center flow $\theta(k)$. The $C_{2z}\mathcal{T}$ symmetry of the Wilson loop $\mathcal{W}(k)$ requires the anti-commutation relation between $C_{2z}\mathcal{T}$ and $H_W$, as it flips the Wannier center\cite{bouhon2019wilson}. Now we consider the expansion of the Wilson Hamiltonian around the red crossing point at $1/2$ in Fig.\ref{fig:phase no I}c, which takes the form $H_W(\delta k)=\delta k\sigma_z$, were $\delta k$ is the momentum away from the crossing point. The above Wilson Hamiltonian anti-commutes with $C_{2z}\mathcal{T}=\sigma_x K$ where $K$ is complex conjugation and we have used the fact that one should define the Wannier center flow $\theta$ module 1 so that $\theta=1/2$ is equivalent to $\theta=-1/2$. Now we will show that $C_{2z}\mathcal{T}$ symmetry can stabilize the crossing at $\theta=1/2$. We consider certain perturbation Hamiltonian $\delta H_W =\sum_{i=x,y,z}\delta_i \sigma_i$. The anti-commutation relation between $C_{2z}\mathcal{T}$ symmetry and $\delta H_W$ requires $\delta_x=\delta_y=0$. Thus, one can see that the remaining $\delta_z$ term can only shift the crossing point to $k=-\delta_z$ for the Wilson Hamiltonian $H_W+\delta H_W$. Thus, we prove the local stability of the crossing at $\theta=1/2$.

\begin{figure}
\centering
\includegraphics[width=1\textwidth]{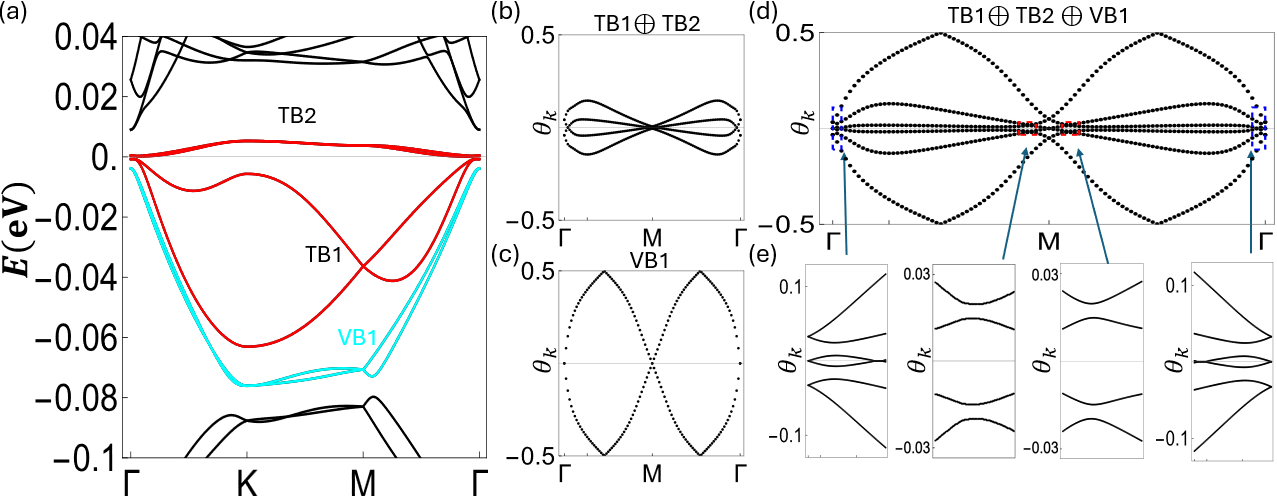}
\caption{\label{fig:fragile test} (a) Band dispertion, when 2 trivial bands highlighted by Red are included. The irrep of VB1 is $(\bar\Gamma_9,\bar K_6)$, the irrep of TB1 is  $(\bar\Gamma_7,\bar K_4\oplus\bar K_5)$, and  the irrep of TB1 is  $(\bar\Gamma_8,\bar K_6)$  (b) Wannier center flow for two 2 trivial bands (c) Wannier center flow for VB1. (d)  Wannier center flow for VB1 and two trivial bands. (e) Zoom in Wannier center flow in the dashed box in d. By including two trivial bands, we can gap out the Wannier center flow  }

\end{figure}

The double winding of $\theta(k)$ indicates that the VB1 in the region B cannot be adiabatically connected to the atomic limit so that the VB1 is non-trivial. Next we will demonstrate that the VB1 carries fragile topology by trivializing the Wannier center flow of the VB1 with adding additional trivial bands. 
The band representation of the VB1 is $(\bar\Gamma_9,\bar K_6,\bar M_5)$. By inspecting the EBR of the space group $P6mm$ \cite{aroyo2011crystallography,aroyo2006bilbao,aroyo2006bilbao1}\addCXL{}, we find the band representation of the VB1 can be combined with an EBR $\bar E_2\uparrow G(2)@1a$ with   $(\bar\Gamma_8,\bar K_6,\bar M_5)$ and an EBR $\bar E_3\uparrow G(2)@1a$ with $(\bar\Gamma_7,\bar K_4\oplus\bar K_5,\bar M_5)$ to make an EBR $\bar E_3\uparrow G(6)@3c$ with $(\bar\Gamma_9 \oplus\bar\Gamma_7\oplus \bar\Gamma_8,\bar K_4\oplus\bar K_5\oplus 2\bar K_6,3\bar M_5  )$. Therefore, we can consider the following Hamiltonian

\begin{eqnarray}\label{eq: Moire BHZ fragile}
&& H_{combine}(k) =\left( \begin{array}{ccc}  H(m,\Delta_1,\lambda)&0&0\\ 
  0& H_T&0\\
0 & 0&\epsilon_2+H(m',\Delta'_1,\lambda')
 \end{array} \right)+H_{int}(k)
 \\
 && H_{int}(k) =\left( \begin{array}{ccc}  0&H_{I1}(k)&H_{I2}(k)\\ 
  H^\dagger_{I1}(k)&0&0\\
H^\dagger_{I2}(k) & 0&0
 \end{array} \right)
 \\
 && H_{T}=(\epsilon_1+\alpha k^2+\Delta''_1 \sum_{\bs g}  e^{i \bs g \cdot \bs r}) I \\
 && H_{I1}(k) =V^0_{int}\left( \begin{array}{cc} 0& (k_x+i k_y)^2 \\ 
i(k_x+i k_y)^2 & (k_x+i k_y) \\
(k_x-i k_y)^2  & 0\\
 (-k_x+i k_y)  & i(k_x-i k_y)^2 
 \end{array} \right) \\ 
 && H_{I2}(k) =V^1_{int} I,
\end{eqnarray}
where $I$ represents a 4-by-4 identity matrix. $H(m,\Delta,\lambda)$ is Moir\'e BHZ model in Eq.(\ref{eq: Moire BHZ}) with the gap parameter $m$, the moir\'e potential $\Delta_1$ and the Rashba SOC parameter $\lambda$. 
The form of the coupling Hamiltonian $H_{int}$ is derived from the requirements of preserving the $\mathcal{M}_y$, $\mathcal{C}_{6z}$ and $\mathcal{T}$ symmetries. Here the parameters are chosen to make the VB1 of the Hamiltonian $H(m,\Delta,\lambda)$ in the region B of the phase diagram in Fig.3c of the main text, and other parameters are chosen as $\epsilon_1$=0.015eV, $\epsilon_2$=-0.05eV, $\alpha$=0.1$nm^2\cdot $eV, $V^0_{int}$=0.1$nm\cdot $eV,  $V^1_{int}$=0.004eV, $\Delta''_1$=-0.01eV, $\Delta'_1$=-0.06eV, $m'$=0.1eV and $\lambda'$=0.4$nm\cdot $eV.    The TB1 minibands (See Fig.\ref{fig:fragile test}a) from the Hamiltonian $H(m',\Delta',\lambda')$ to possess the band representation $(\bar\Gamma_7,\bar K_4\oplus\bar K_5)$, and the TB2 minibands (See Fig.\ref{fig:fragile test}a) from the Hamiltonian $H_T$ belonging to the band representation $(\bar\Gamma_8,\bar K_6)$. The full band dispersion of the model $H_{combine}$ is shown in the Fig.\ref{fig:fragile test}a. 
The $\theta(k)$ for the TB1 and TB2 in Fig.\ref{fig:fragile test}b exhibits a trivial feature. The $\theta(k)$ for the VB1 of the Hamiltonian $H_{combine}$ still reveals a double winding feature in the  Fig.\ref{fig:fragile test}c, and thus the band topology of the VB1 remains the same. By including two additional trivial bands of TB1 and TB2 in the calculation of Wannier center flow, as shown in the Fig.\ref{fig:fragile test}d and e, we find the overall Wannier center flow $\theta(k)$ for all three minibands, the VB1, TB1 and TB2, are fully gapped and become trivial. Therefore, we demonstrate that the doubling winding of the VB1 in region B of Fig.3c of the main text can be gapped out by introducing trivial bands, thus belonging to the fragile topology. 


\end{document}